\documentclass[10pt,prb,aps,twocolumn,superscriptaddress,showpacs,amsfonts,amsmath,amssymb,showkeys,floatfix]{revtex4-1}

\usepackage{bm}
\usepackage{pstricks}
\usepackage{setspace}
\usepackage{fdsymbol}
\usepackage{amssymb}
\usepackage{graphicx}% Include figure files

\begin{document}
\title{Magnetic ordering of random dense packings of freely rotating  dipoles.   
}

\date{\today}
\author{Juan J. Alonso}
\email[e-mail address: ] {jjalonso@uma.es}
\affiliation{F\'{\i}sica Aplicada I, Universidad de M\'alaga, 29071 M\'alaga, Spain}
\affiliation{Instituto Carlos I de F\'{\i}sica Te\'orica y Computacional,  Universidad de M\'alaga, 29071 M\' alaga, Spain}
\author{B. All\'es}
\email[E-mail address: ] {alles@pi.infn.it}
\affiliation{INFN--Sezione di Pisa, Largo Pontecorvo 3, 56127 Pisa, Italy}
\author{V. Russier}
\email[E-mail address: ] {russier@icmpe.cnrs.fr}
\affiliation{ICMPE, UMR 7182 CNRS and UPE 2-8 rue Henri Dunant 94320 Thiais, France.}

\date{\today}
%\pacs{75.10.Nr, 75.10.Hk, 75.40.Cx, 75.50.Lk}

\begin{abstract}
We study random dense packings of Heisenberg dipoles by numerical simulation.
The dipoles are at the centers of identical spheres that occupy fixed random positions in space and fill a fraction $\Phi$ of the spatial volume.
The parameter $\Phi$ ranges from rather low values, typical of amorphous ensembles, to the maximum $\Phi$=0.64 that occurs in
the random-close-packed limit. We assume that the dipoles can freely rotate and have no local anisotropies. 
As well as the usual thermodynamical variables, the physics of such systems depends on $\Phi$.
Concretely, we explore the magnetic ordering of these systems in order to depict the phase diagram in the temperature-$\Phi$ plane.
For $\Phi \gtrsim0.49$ we find quasi-long-range ferromagnetic order coexisting with strong long-range spin-glass order.
For $\Phi \lesssim0.49$ the ferromagnetic order disappears giving way to a spin-glass phase similar to the ones found
for Ising dipolar systems with strong frozen disorder.
\end{abstract} 
%\submitted
\maketitle

\section{INTRODUCTION}
\label{intro}

The problem of identifying the magnetic ordering induced by a dipolar interaction has been attracting a renewed interest.\cite{fiorani, sawako1}
This is due to the surge of innovative materials built by assembling magnetic nanoparticles (NP) into dense packings. The interest of such materials lies in the perspective of a plethora
of applications that they may offer, in particular in nanomedicine, nanofluids, or in data storage.\cite{pang, np, bedanta}

NP are synthesized with Cobalt, Iron or Iron Oxydes, then coated with layers of non-magnetic material, and finally laid into monodisperse systems.\cite{nano} 
NP a few tens of nanometers wide behave like permanent magnets with magnetic moments ranging between
$10^{3}$ and $10^{5}$ Bohr magnetons. These NP often exhibit anisotropy energy barriers $E_{a}$ that trigger the ordering along local easy axes.\cite{neel} 
However, the dipolar interaction energies $E_{dd}$ can become quite large in dense
packings, even larger than $E_a$. When this occurs, dipolar induced magnetic order is observed at temperatures that are low but still above the blocking
temperature $k_BT_{b}\simeq E_{a}/30$, where $k_B$ is the Boltzmann constant. This is in contrast to the super-paramagnetism that is observed 
in not very dense systems.\cite{superpara} 

Luttinger and Tisza showed that freely rotating dipoles
placed in face-centered cubic (FCC) or body-centered cubic (BCC) networks possess ground states with ferromagnetic (FM) order.
When they are placed on a simple cubic (SC) lattice, antiferromagnetic (AF) order is found instead.\cite{lutti} 
These results are supported by numerical Monte Carlo (MC) simulations.\cite{bouchaud, silvano}
Recently the necessary technology for synthesizing NP has been
developed allowing to obtain crystalline orderings of NP, thus opening the possibility of investigating by empirical means the FM and AF orders in such supercrystals.\cite{sc1,sc2} 

However, a certain structural disorder, be it positional or orientational, is often present in dense systems. 
The magnetic order strongly depends on the
relative positions of the NP and, due to the specific anisotropy in the dipolar interaction, on the relative orientations of the
easy axes existing in presence of local anisotropies. Both types of disorder can spoil the large-order behavior giving rise to spin-glass (SG) behavior.
This phenomenon has been experimentally observed in frozen ferrofluids,\cite{ferrofluids, morup}
and in random dense packings (RDP) of dipolar spheres with volume fractions $\Phi \approx 0.64$
obtained by pressing powders.\cite{powders, toro1}

The role played by the degree of orientational disorder, called texturation, in the magnetic order has been studied by MC simulations both in FCC lattices
and in RDP.\cite{jpcm17, alonso19, russier20}
  In particular, the phase diagram of non-textured FCC systems has been obtained as a function of $E_{a}/E_{dd}$,\cite{enviado} 
% completar cita antes de enviar a publicar  
where the ratio $E_{a}/E_{dd}$ is an estimate the degree of disorder in such non-textured lattices.

On the other hand, the relevance of positional disorder is a controversial issue, far from being completely understood. This is the subject of the present paper.

Although strictly speaking there cannot be single domains of NP without local anisotropy, we study
the effect of the positional disorder on the magnetic ordering in the limiting case of Heisenberg
dipoles free of anisotropy. This is because we wish to understand the consequences of pure positional disorder, without intereferences from the anisotropy disorder.
Numerical simulations show that dipolar spheres moving in a non-frozen fluid
exhibit long-range nematic order even for volume fractions as low as $\Phi=0.42$.\cite{weis, weis2} Such systems
develop spatial correlations at low temperatures that do not exist in the case of frozen ferrofluids. Long-range order has been observed for the former.
Then, the key question is: can long-range order appear in systems with frozen positional disorder without fine tuned positional correlations? MC simulations of freely rotating
dipoles (i.e., Heisenberg dipoles with $E_{a}=0$) in fluid-like amorphous frozen configurations with $\Phi=0.42$ show no trace of strong FM order in the thermodynamic limit.
They showed only signatures of orientational freezing at low temperature.\cite{ayton1,ayton2}
Zhang and Widom considered a mean-field  approximation for systems of frozen dipolar hard spheres 
randomly distributed ocuppying a fraction $\Phi$ of the volume. By using the approximation $g(r) = 1$ for the radial
distribution function regardless of the value of $\Phi$,  they found long-range FM order for $\Phi \ge 0.295$ in contrast to the results from simulations.\cite{zhangwidom}
Recently, numerical evidence of SG order has been found in strongly diluted systems 
of Heisenberg dipoles in SC lattices.\cite{stasiak, zhang-dilu}

In this paper we study by MC simulations the magnetic order in RDP made up of Heisenberg NP with $\Phi$ ranging from low values to the maximum $\Phi=0.64$ (this is the number taken by this parameter when the system is a
random-close-packed (RCP) ensemble.)\cite{torquato} The dipoles will be free to rotate, but their positions, albeit randomly distributed, will be regarded as
fixed. Precisely, the only allowed structural disorder will be this randomness in the NP positions. We want to study if this disorder is able to spoil the FM arrangement to produce a SG phase. Concretely,
we will investigate whether short range spatial correlations in RDP (see Fig.~\ref{gR}), can allow some type of FM order for $0.42 < \Phi \leq 0.64$.
The occupied fraction $\Phi$ of the volume will be used to rate the degree of disorder and, in fact we will obtain
a phase diagram showing the distribution of equilibrium phases in the temperature-$\Phi$ plane. We will also analyse the nature of the several phases, by using data
taken from measurements of the magnetization, the SG overlap parameter,\cite{ea} and related fluctuations.

The paper is organized as follows. In Sec.~\ref{mm} we will introduce the model, describe the MC algorithm and list the
definitions of the several observables that shall be measured. We will present and discuss the outputs of those measurements in
Sec.~\ref{results}. In Sec.~\ref{results} we also analyse the degree of disorder as a function of $\Phi$. A summary of the results obtained
in the paper will be given in Sec.~\ref{conclusion}, together with a few concluding remarks.

\section{MODEL AND SIMULATION DETAILS}
\label{mm}

\subsection{Model}
\label{models}

We will consider RDP composed by $N$ identical NP that behave as single magnetic Heisenberg dipoles. The NP will be labelled with an index $i=1,\dots,N$.
Each NP is a sphere of diameter $d$. The magnetic moment of the $i$--th NP 
will be denoted by $\vec{\mu}_{i}=\mu\widehat{\sigma}_{i}$ 
where $\widehat{\sigma}_i$ is a unit norm direction.
We will be concerned only with the dipole-dipole interactions between NP. Moreover, no local anisotropy will be assumed in such a way that each magnetic moment can rotate  freely.

The Hamiltonian reads
\begin{equation}
{\cal H}= \sum_{ <i,j>}  \varepsilon_d\left( \frac {d}{r_{ij}} \right) ^{3}
\Big( \widehat{\sigma}_i \cdot \widehat{\sigma}_j-\frac {3(\widehat{\sigma}_i\cdot \vec{r}_{ij})( \widehat{\sigma}_j\cdot \vec{r}_{ij})} {r_{ij}^2}\Big)\;,
\end{equation}
where $\varepsilon_d =\mu_{0}\mu^2/(4 \pi d^{3})$ is an energy and $\mu_0$ the magnetic permeability in vacuum.
$\vec{r}_{ij}$ is the vector position of dipole $j$ viewed from dipole $i$, and $r_{ij}=\Vert\vec{r}_{ij}\Vert$,
The summation runs over all pairs $i,j$ of different NP. The positions of the spherical NPs are frozen.

Such arrangements can be obtained with the Lubachevsky-Stillinger (LS) algorithm.\cite{ls, donev}
It consists in the following steps. Firstly, $N$ very small spheres are
placed at random by try and error in a cube of edge $L$. Secondly, the spheres are allowed to move and collide
as hard-spheres while growing in size. During all this process, periodic boundary conditions are
assumed. Furthermore, the growing rate is chosen to be sufficiently large in order to permit the sample to get eventually
stuck in a RCP structure at the maximum possible volume fraction $\Phi=0.64$ before reaching any equilibrium configuration.\cite{torquato, donev}   
Configurations with smaller values of $\Phi$ can be achieved by using the same recipe and stopping
when the desired value of  $\Phi$ has been attained within a $2 \text{\textperthousand}$ precision.
Note that when the LS procedure stops, the spheres have reached a diameter $d=L ({{6\Phi}/{N \pi}})^{1/3}$.

\begin{figure}[!t]
\begin{center}
\includegraphics*[width=65mm]{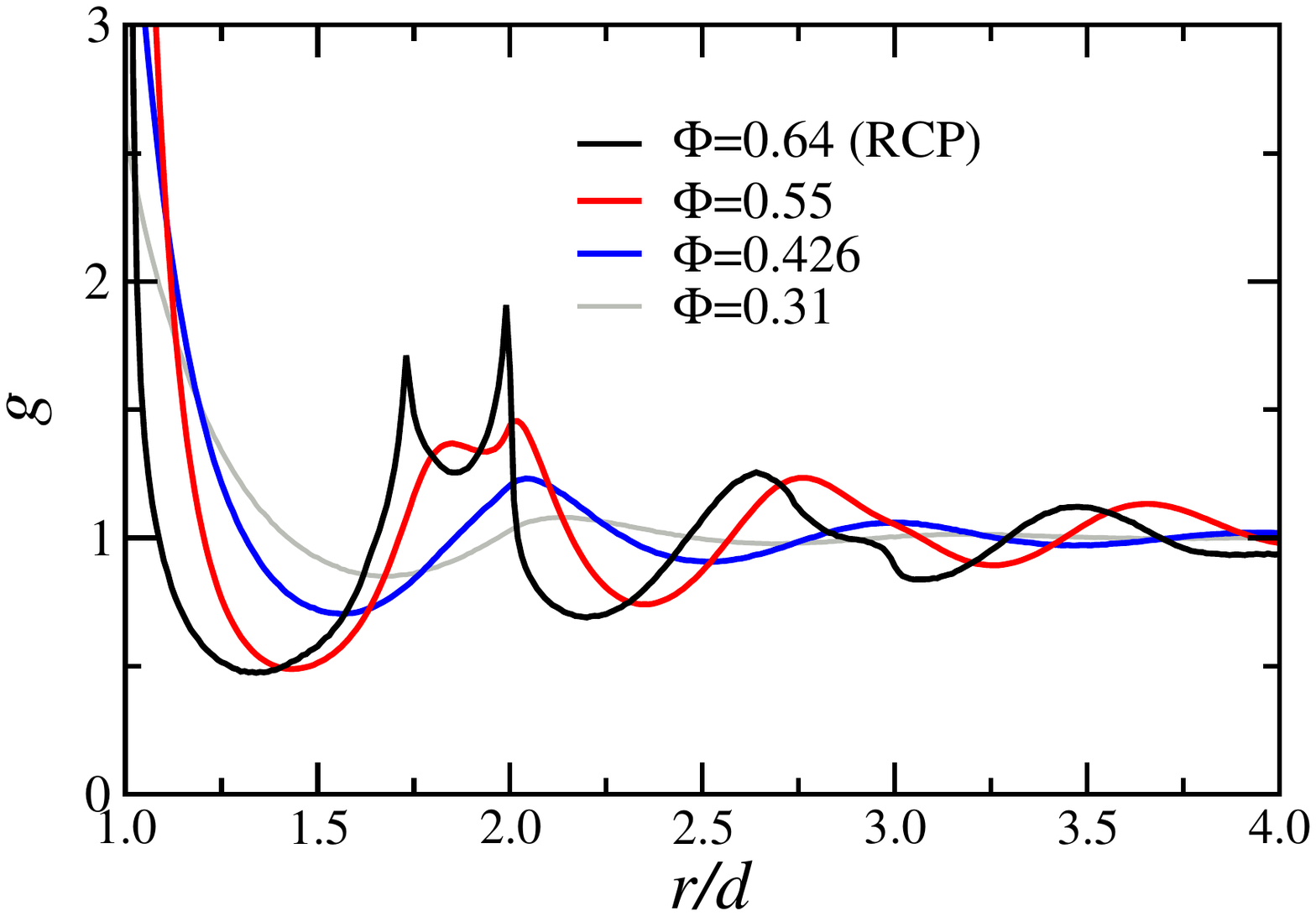}
\caption{(Color online) Radial distribution function 
$g(r)$ for disordered dense packings of $N=8000$ particles obtained with the LS algorithm
for several values of the disorder parameter $\Phi$. The positions of the peaks of the double horn in the curve for RCP 
coincide precisely with $r/d=\sqrt{3}$ and $2$, 
the $3^{th}$ and $4^{th}$ nearest neighbor distances in FCC lattices respectively
 ($\sqrt{2}d$ in FCC  is the lateral size of the  face-centered cubes).
A lingering signature of the double peak persists at $\Phi=0.55$.
The absence of peaks at $r/d=\sqrt{2}$ and $\sqrt{5}$ indicates that there is not crystalline order in the packings.\cite{torquato}}
\label{gR}
\end{center}
\end{figure}

The out-of-equilibrium random packings of spheres produced by the LS method mimic empirical packings,
namely they are similar to the samples obtained by raw compression
of powders of NPs, or to those achieved by suddenly freezing colloidal suspensions of NP's.
For densities below the freezing point ($\Phi = 0.49$),\cite{santos} we find that 
our radial distribution function $g(r)$ is very close to that of the hard sphere fluid at equilibrium. 
For $\Phi \gtrsim 0.49$, on the other hand, the LS method
provides  configurations that do not show significant crystal nucleation and are near to the metastable branch 
whose ending point is the RCP limit.\cite{torquato} Note that, contrary to the case of ferrofluids, here there are no 
spatial correlations other than those due to steric constraints.

In Fig.~\ref{gR} we plot the radial distribution function for four values of $\Phi$ in ensembles with $N=8000$ obtained with the LS method.
The double horn shape found in the RCP case indicates the existence well-tuned short-range spatial correlations.
Our aim is to investigate whether such random packings may develop some kind of dipolar FM order for
dense enough systems as it is the case for dipolar fluids or for systems of dipoles placed on the sites of FCC lattices for which strong long-range FM order is known to appear. 

In what follows, distances and temperatures $T$ will be given in units of $d$ and $\varepsilon_d/k_{B}$ respectively.

\subsection{Method}

\begin{figure}[!t]
\begin{center}
\includegraphics*[width=80mm]{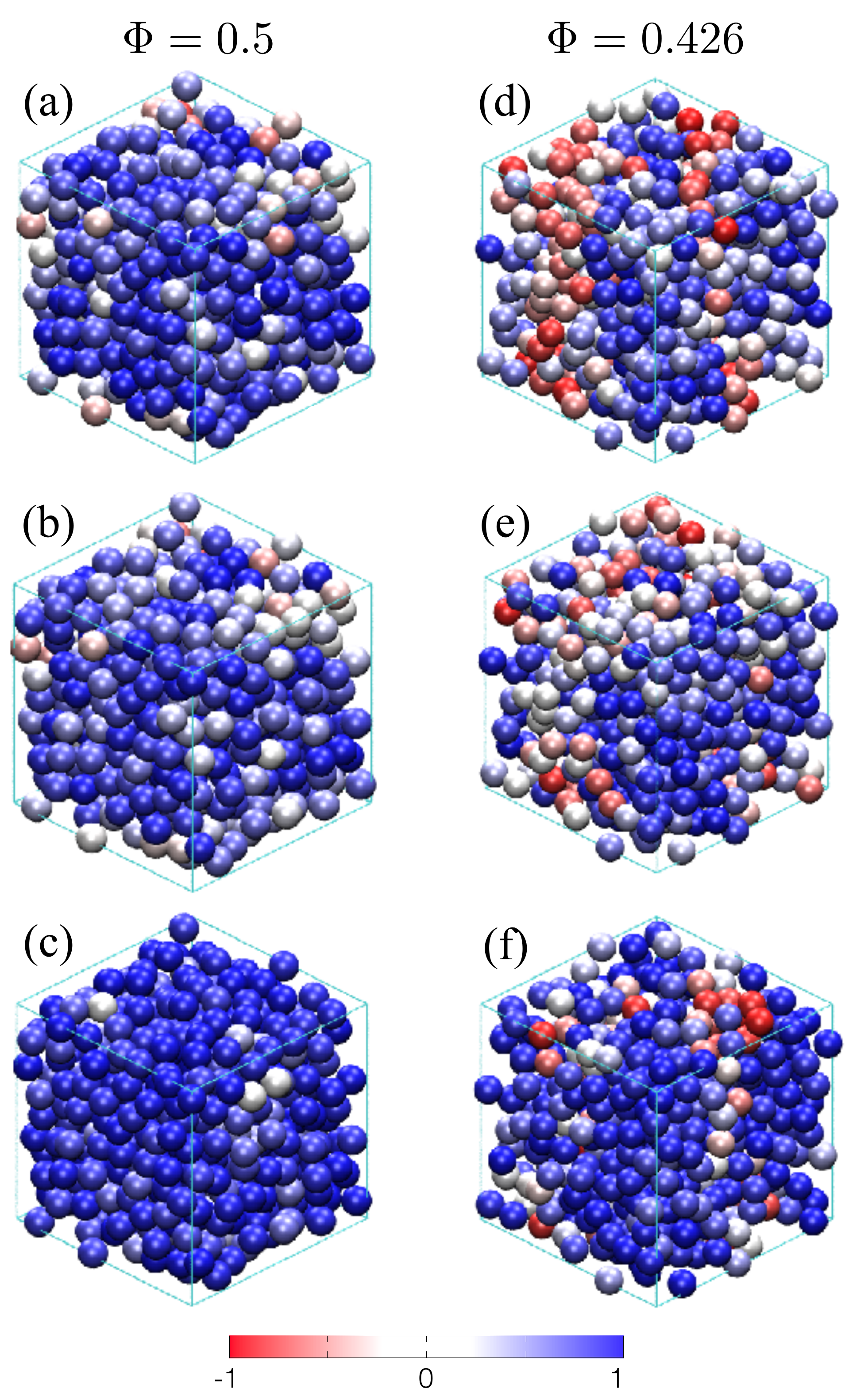}
\caption{(Color online) Pictures (a) and (b) show two independent 
configurations of a given sample with $N=512$ particles with $\Phi=0.5$ at temperature $T=0.1$.
  The position of the spheres are frozen. The hue assigned to each sphere $i$ gives a measure of the degree of correlation $\widehat{\sigma}_{i}^{(A)} \cdot  \widehat{\lambda}$ where $A=a,b$,
  between the magnetic moment $ \widehat{\sigma}_{i}$ and the nematic director vector $\widehat{\lambda}$ of the configuration.
  Picture (c) represents the overlap between the configurations (a) and (b). In this case the color of spheres give an idea of the parallelism defined by the product
  $ \widehat{\sigma}_{i}^{\rm (a)} \cdot  \widehat{\sigma}_{i}^{\rm (b)}$. The analogous pictures (d), (e), (f) exhibit the same properties for two configurations of a sample at $\Phi=0.426$.
    The color scale in the bottom exhibits the correspondence between hue and degree of alignment between the vectors in the scalar products (from +1 when are parallel, to $-1$ when are antiparallel).
}
\label{portrait}
\end{center}
\end{figure}

Since a certain SG-like behavior is expected to show up, at least for small values of $\Phi$, we will employ familiar SG notations.
Concretely, any system of NP with a specific realization of randomness $\Phi$, with the positions of all NP fixed, will be called {\it sample} and denoted by $\cal J$.
In Figs.~\ref{portrait}(a) and (d) two samples are shown, one for $\Phi=0.5$ and the other for $\Phi=0.426$. The positions of the $N$ NPs are fixed and
only the magnetic moments $\widehat{\sigma}_{i} $ participate in the dynamics. We will call {\it configuration} any list of $N$ unit vectors
$\{\widehat{\sigma}_{i}\}_{i=1,\dots,N} $ in any given sample.

For a given temperature $T$ and a given sample ${\cal J}$, the MC simulation provides a set of thermally distributed configurations. The average of any physical
quantity calculated for each element of this set and averaged over all the set, gives an estimate of that quantity. Nevertheless, in order to get physical results ready
to be compared with experimental measurements, a second average, this time over $N_s$ independent samples at the same temperature $T$, is performed.
The need of this second average is particularly important for small $\Phi$, where large sample-to-sample fluctuations are expected to occur. 
The numbers of samples $N_{s}$ for the values of $N$ and $\Phi$ used in our simulations are shown in Table~I.
From this table it is evident that we do not make $N_s \propto 1/N$  for small values of $\Phi$, due the well-known non-self-averaging property of SG systems.

The samples are expected to exhibit strong frustration and rough free energy landscapes, at least for small values of $\Phi$. In principle this property can heavily slow down the simulation.
Then, with the purpose of obtaining truly thermalized
sets of configurations in reasonable computer times, we resorted to the tempered Monte Carlo (TMC) algorithm.\cite{tempered}
It consists in running in parallel $n$ identical replicas of each sample at slightly different temperatures within an interval $[ T_{\rm min},T_{\rm max}]$. The $n$ temperatures are separated by
an amount $\Delta$, so they are
  $T_{\rm min}$, $T_{\rm min}+\Delta$, $T_{\rm min}+2\Delta$, \dots $T_{\rm min}+(n-2)\Delta$, $T_{\rm min}+(n-1)\Delta\equiv T_{\rm max}$.
Each of these values and its neighbor are called neighbor temperatures. Every replica is let to evolve independently by 10~MC sweeps of the usual heat-bath (HB) algorithm.\cite{heatbath}
Then, the HB algorithm is stopped to allow the replicas from neighbor temperatures to be exchanged while respecting detailed balance.\cite{tempered}
 Once all permitted exchanges have been performed, the process is reinitiated with another 10~sweeps of HB.
The values of $\Delta$ are selected in such a way that roughly 30\% of the exchanges be accepted. The TMC parameters are given in the caption of Table~\ref{table1}.

\begin{table}[!t]
\begin{tabular}{p{1.1cm} p{1.0cm } p{1.0cm } p{1.0cm } p{1.0cm } p{1.0cm }}
\multicolumn{6}{c}
{$\Phi=0.64$ (RCP)}\\
 \hline
$N$ & $64$ &$125$ &$216$ & $512$ &$1000$  \\
$N_s$ & $4500$ & $3500$ & $2000$ &$1000$ &$1000$  \\
\hline
\multicolumn{6}{c}
{$\Phi=0.55$}\\
 \hline
$N$ & $64$ &$125$ &$216$ & $512$ &$1000$  \\
$N_s$ & $8000$ & $3500$ & $2000$ &$2000$ &$1600$  \\
\hline
\multicolumn{6}{c}
{$\Phi=0.5$}\\
 \hline
 $N$ & $64$ &$125$ &$216$ & $512$ \\
$N_s$ & $8000$ & $8000$ & $7500$ &$4000$ &  \\
\hline
\multicolumn{6}{c}
{$\Phi=0.465$}\\
 \hline
$N$ & $64$ &$125$ &$216$ & $512$ &  \\
$N_s$ & $8000$ & $6000$ & $6000$ &$5000$ &  \\  
\hline
\multicolumn{6}{c}
{$\Phi=0.426$}\\
 \hline
$N$ & $64$ &$125$ &$216$ & $512$ &  \\
$N_s$ & $8000$ & $8000$ & $7800$ &$7000$ &  \\
\hline
\multicolumn{6}{c}
{$\Phi=0.31$ }\\
 \hline
$N$ & $64$ &$125$ &$216$ & $512$ &    \\
$N_s$ & $8000$ & $8000$ & $8300$ &$7000$ &   \\
\hline
\end{tabular}
\caption{The parameters utilized in the TMC simulations. $\Phi$ is the volume fraction, $N$ the number of dipoles, $N_{s}$ the number of samples.
  We used a step $\Delta=0.025$ for temperatures $T\le 0.6$ and $\Delta=0.05$ for $T > 0.6$. The highest temperature was $T_{\rm max}=1.1$.
  The lowest temperatures for $\Phi \ge 0.426$ was $T_{\rm min}=0.1$ for $N \le 512$, and $T_{\rm min}=0.175$ for $N=1000$; while for $\Phi =0.31$
  it was $T_{\rm min}=0.05$ for $N\le216$, and $T_{\rm min}=0.1$ for $N=512$.
  The number $t_0$ of initial MC sweeps for equilibration was at least $t_{0}=10^{6}$, and the measurements were taken within the interval $[t_{0}, 2t_{0}]$.
 }
\label{table1}
\end{table}

Periodic boundary conditions were used in the simulations. Any dipole $i$ interacts with the dipoles within a cube $L\times L\times L$ centered at $i$.
The long-range dipolar-dipolar interaction was treated by Ewald's sums.\cite{ewald,holm} In these sums we split the computation of the dipolar fields into a real
space sum with a cutoff $r_{c}=L/2$ and a sum in the reciprocal space with a cutoff $k_{c}$ by screening each dipole with a distribution with standard deviation
$\alpha$. We have used $\alpha=4/L$ and $k_{c}=10(2\pi/L)$.\cite{holm} Given that we focus the study on the search of FM order, any possible shape dependent
demagnetizing effect was avoided by using the so-called conductive external conditions (i.e. using a surrounding permeability $\mu^\prime=\infty$.)\cite{weis,allen}

The thermal equilibration times $t_{0}$ were estimated after examining the plateaux for large time $t$ of the overlap parameter $q$
(see next Section) starting from different
initial configurations  as described at length in Refs.\cite{jpcm17, PADdilu2} We also verify the symmetry in the thermal distributions of magnetization
and the SG overlap parameter under the global inversion $\{\widehat{\sigma}_{i}\} \to  \{-\widehat{\sigma}_{i}\}$  as an additional check that all samples are well equilibrated.\cite{jpcm17}
We used the first $t_{0}$ MC sweeps to equilibrate the samples and all thermal averages were extracted in the interval $[t_0,2 t_0]$. As mentioned above, a second average over
$N_{s}$ samples is performed in order to obtain physical results.  These double average will be indicated by angular brackets $\langle \cdots\rangle$.

\subsection{Observables}
\label{meas}

Our aim is to investigate the nature of the low temperature ordered phases and determine the
transition temperature between theses phases and the high temperature paramagnetic (PM) phase as a function of the volume fraction $\Phi$. 
In this subsection we introduce the physical quantities that we have deemed adequate for that purpose.

To explore the possible existence of nematic order we have extracted
the eigenvector with largest eigenvalue $P_{2}$ of the tensor
${\pmb{ \mathbb Q}} \equiv  \frac{1}{2N} \sum_{i} (3 \widehat{\sigma}_i \otimes \widehat{\sigma}_i  -     \pmb{\mathbb{I}} )$.
Once normalized, this eigenvector is called nematic director, $\widehat{\lambda}$.\cite{weis,allen} $P_{2}$ is in fact the nematic order parameter
\begin{equation} 
P_2\equiv \frac{1}{2N} \sum_i  [ 3(\widehat{\sigma}_{i} \cdot \widehat{\lambda})^{2} -1] \;.  
\label{nema}
\end{equation}
The double average of this quantity gives the degree of global alignment of all dipoles along the director $\widehat{\lambda}$.

The magnetization vector is defined as $\vec{m} \equiv (1/N) \sum_i \widehat{\sigma}_i$. Instead of $\Vert{\vec{m}}\Vert$ we use as FM order parameter the projection of $\vec{m}$
\begin{equation} 
m_{\lambda} \equiv  \frac{1}{N} \sum_i  ( \widehat{\sigma}_{i} \cdot \widehat{\lambda} )  \;,  
\label{ml}
\end{equation}
along $\widehat{\lambda}$ . Nonetheless, according to our simulations both quantities provide qualitatively the same results. 

We have also computed the moments $m_{p}=\langle |m_{\lambda}|^{p}\rangle$ for $p=1,2,4$. These moments allow to calculate the magnetic susceptibility
\begin{equation}
\chi_{m}\equiv {N\over{k_{B}T}}(m_{2}-m_{1}^{2})\;.
\label{suscZ}
\end{equation}
and the Binder cumulant
\begin{equation}
B_{m}\equiv{1\over 2} (3-{m_{4} \over m_{2}^{2} })\;.
\label{Bm}
\end{equation}
The dimensionless quantity $B_m$ will turn out useful for locating the PM-FM transition temperature.

The specific heat $c_{v}$ is obtained from the fluctuations of the energy $e\equiv\langle {\cal H}\rangle/N$.

For investigating the SG order we use  the overlap parameter between replicas (1) and (2) of a given sample
\begin{equation} 
q_\equiv \frac{1}{N} \sum_i  \widehat{\sigma}_{i}^{(1)} \cdot \widehat{\sigma}_{i}^{(2)}\;,  
\label{q}
\end{equation} 
  instead of the more familiar tensorial quantities
  \begin{equation} 
  q_{3d} \equiv  \sum_{\alpha,\beta=1}^{3} |q_{\alpha\beta}|^{2}, \quad\textrm{with}\quad
  q_{\alpha\beta}\equiv \frac{1}{N} \sum_i  {\sigma}_{i \alpha}^{(1)} {\sigma}_{i\beta}^{(2)}\;,
  \label{q3d}
  \end{equation}
  often used when dealing with Heisenberg spins. ${\sigma}_{i \alpha}^{(A)}$ in (\ref{q}) is the $\alpha$ component of the unit vector $\hat{\sigma}^{(A)}_i$ of the $A$-th replica, ($A=1,2$).
  The reason why we decline using $q_{3d}$ is that $q_{3d}$ is invariant under global rotations of all dipoles in the configuration, while
  we prefer to keep track of any possible rotation experienced by the nematic director during the simulation.

Similarly to the FM case, we compute the moments  $q_{p}\equiv \langle |q|^{p}\rangle$ for $p=1,2,4$ and calculate the Binder parameter
\begin{equation}
B_{q}\equiv{1\over 2} (3-{q_{4}  \over q_{2}^{2} })\;.
\label{nuevaequationdeTomeu}
\end{equation}

In order to facilitate the identification of the PM-SG transition line we also use the so called SG correlation length,\cite{longi,balle} given by
\begin{equation} 
\xi^2_L\equiv\frac {1 } {4 \sin^2  ( k /2)}  { \left( \frac{q_2} { \langle\mid q(\vec{k})  \mid ^2   \rangle}  -1 \right) }\;,
\label{phi1}
\end{equation}
where $q(\vec{k})$ is
\begin{equation} 
q({\vec{k}})\equiv \frac{1}{N} \sum_i  \psi_{i}~ e^{{\rm i} \vec{k}\cdot \vec{r}_i}\;,
\label{phi2}
\end{equation}
with
$\psi_{i}=\widehat{\sigma}_{i}^{(1)}\cdot\widehat{\sigma}_{i}^{(2)}$,
$\vec{r}_i$ the position of dipole $i$,
$\vec{k} =(2\pi/L,0,0)$ and $k=\Vert\vec{k}\Vert$.
In the PM phase, $\langle\psi_r\psi_0\rangle$ decays in the thermodynamic limit as $\exp(-r/\xi_\infty)$ where
$\xi_\infty$ is the correlation length. At high temperatures, $\xi_L$ in Eq.(\ref{phi1})  provides a good  approximation of $\xi_\infty$.\cite{balle}

We also compute the thermal probability distributions  $p(m_{\lambda})$ and $p(q)$, averaged over all samples.

The errors in the measurements of all averaged quantities were assessed with the mean squared deviations of the sample-to-sample fluctuations. 
In order to minimize these errors, we have enlarged $N_s$ as much as possible within the CPU-time resources available.
The larger the positional disorder is (i.e., the smaller $\Phi$ is), the wilder these fluctuations appear. Also the relaxation times increase with diminishing $\Phi$.
It is for this reason that (i) we were obliged to limit the system sizes
for small $\Phi$ to be no larger than $N=512$ and (ii) systems at temperatures much less
than half the transition temperature were not explored.

\begin{figure}[!b]
\begin{center}
\includegraphics*[width=84mm]{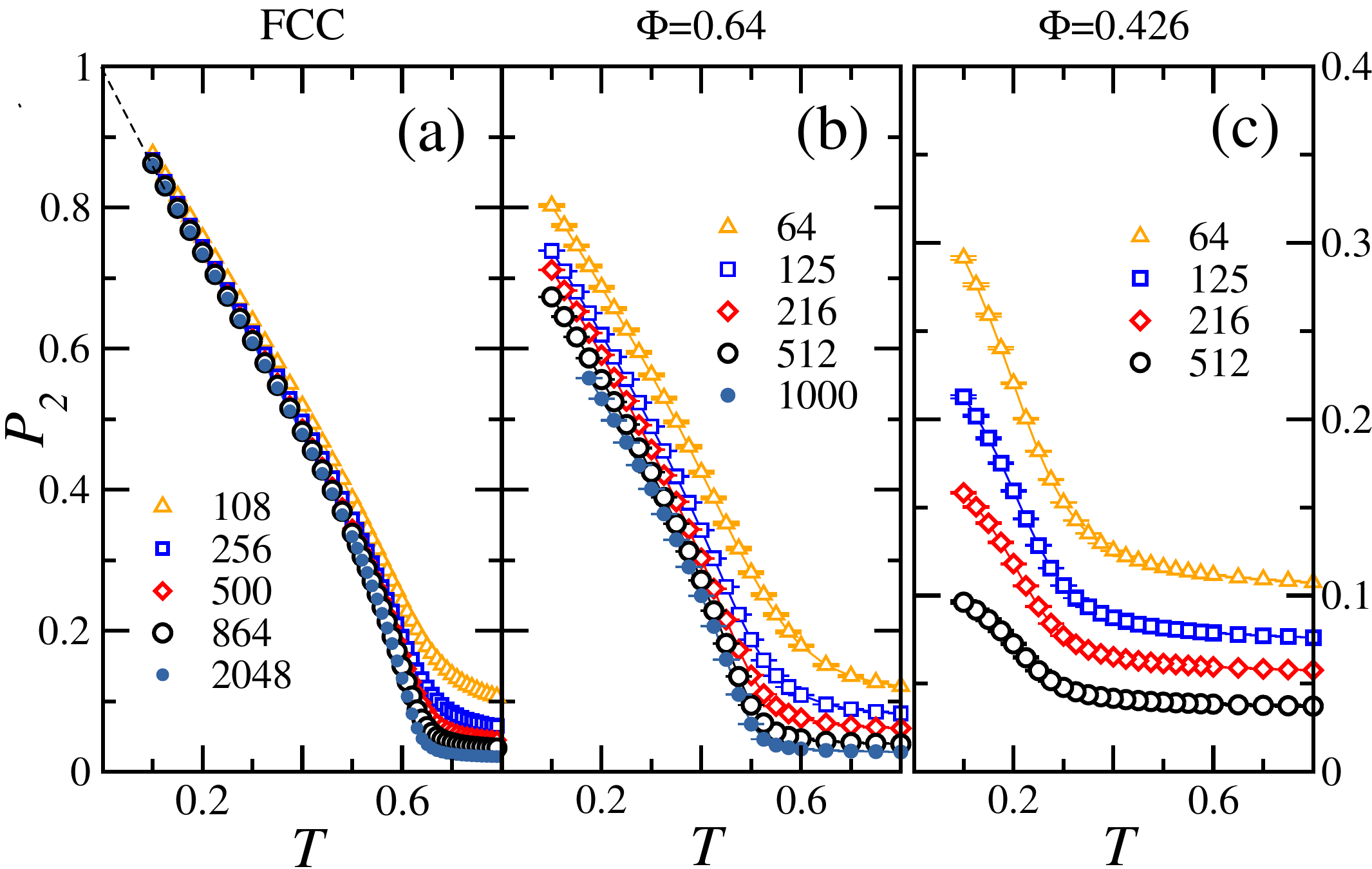}
\caption{
(Color online)
(a) Plots of the nematic order parameter $P_{2}$ versus  $T$ for the FCC lattice and several numbers $N$ of dipoles. 
(b) Same as in (a) for RCP configurations ($\Phi=0.64$). (c) Same as in (b) for $\Phi=0.426$. Lines in all panels are guides to the eye.
}
\label{nematic}
\end{center}
\end{figure}

\section{RESULTS}
\label{results}
A rough estimate of the kind of magnetic order at a given volume fraction $\Phi$ can be grasped by
examining equilibrium configurations at very low temperature for a single sample. Figs.~\ref{portrait}(a) and \ref{portrait}(b)
show two independent configurations, called $(a)$ and $(b)$,
for a sample at $\Phi=0.5$. The hue with which each nanoparticle
has been colored represents the degree $(\widehat{\sigma}^{(A)}_{i} \cdot \widehat{\lambda})$ of alignment between the
dipole associated with the particle and the nematic director ($A=a,b$).
Both configurations exhibit a large magnetic domain with the presense of some non-negligible disorder.
However, the significant overlap between both configurations (see Fig.~\ref{portrait}(c), where now the
hue represents $ (\widehat{\sigma}_{i}^{(a)} \cdot \widehat{\sigma}_{i}^{(b)})$), indicates that that disorder is in reality due to the presence
of SG order. This suggests the existence of partial FM order together with  stronger
SG order at the same time. Configurations with larger $\Phi$ show a similar behavior.

All that is in sharp contrast with the behavior encountered for $\Phi=0.426$. In the configurations of Figs.~\ref{portrait}(d) and (e)
magnetic domains with opposite signs are seen coexisting. However the overlap between the two configurations (see Fig.~\ref{portrait}(f))
is still sizeable and this fact is an indication that SG order dominates any FM order.
Plots of the nematic order parameter $P_{2}$ vs $T$ for different sizes supply additional information about the nature of the phases.
A direct comparison between Figs.\ref{nematic}(a) and (b) reveal a qualitative behavior that differs in the cases of FCC and of RCP (i.e., with $\Phi=0.64$.)
For FCC $P_{2}$ is clearly different from zero and independent of the size at low $T$, as it was to be expected for a dipolar ferromagnet.
Instead, for $\Phi=0.64$ $P_{2}$ decreases when $N$ increases for all $T$.
Plots of $P_{2}$ vs $N$ (not shown) indicate that this trend is algebraic at low temperatures.
Finally, the plots corresponding to $\Phi=0.426$ (see Fig.~\ref{nematic}(c)) evidence absence of nematic order in the thermodynamic limit.

\begin{figure}[!t]
\begin{center}
\includegraphics*[width=84mm]{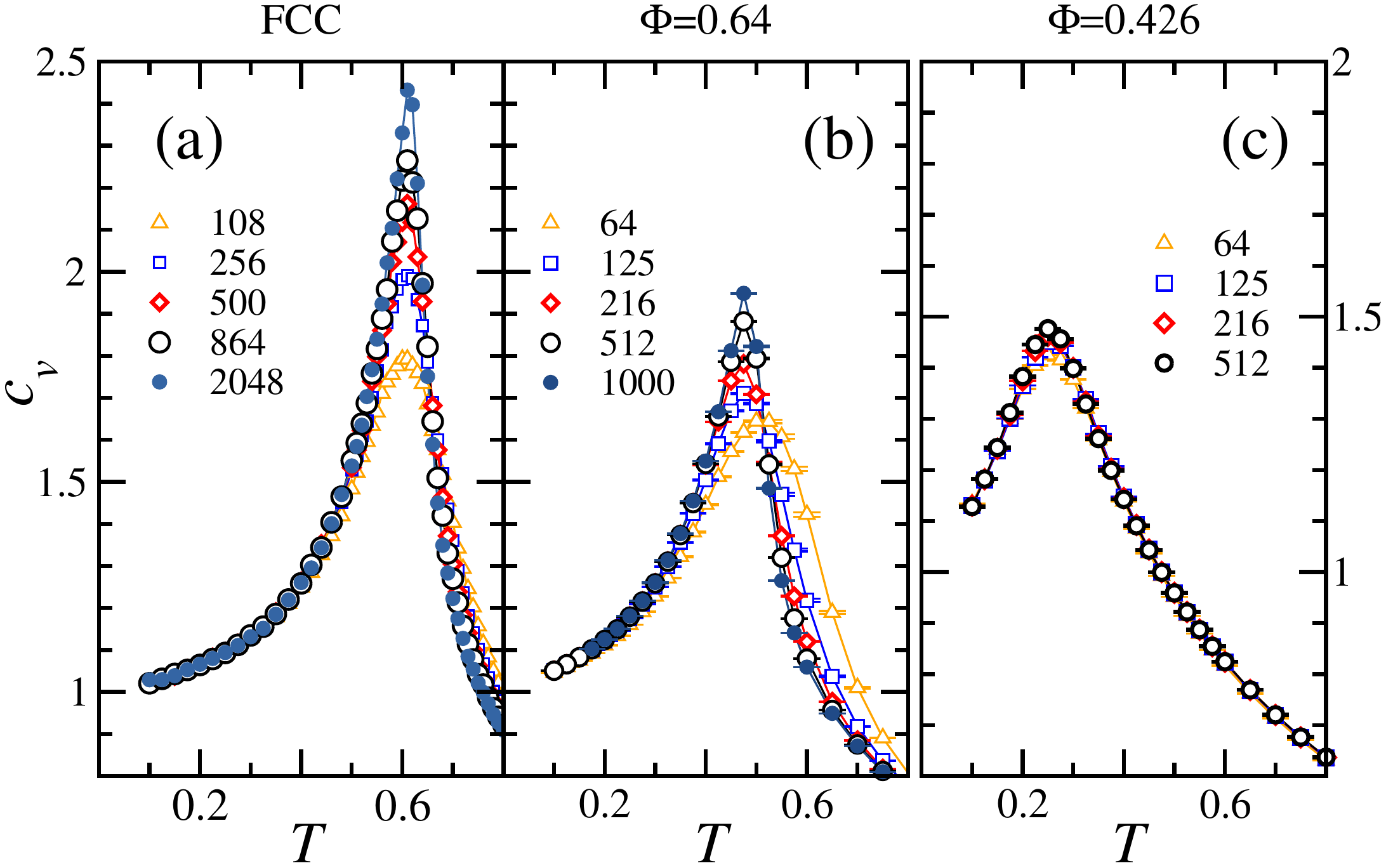}
\caption{ (Color online)
(a) Plots of the specific heat $c_{v}$  versus  $T$ for the FCC lattice and the number of dipoles $N$ indicated in the Figure.
(b) Same as in (a) for RCP configurations ($\Phi=0.64$). (c) Same as in (b) for $\Phi=0.426$. Solid lines in all panels are guides to the eye.}
\label{heat}
\end{center}
\end{figure}

Figs.~\ref{heat}(a) and (b) exhibit plots of the specific heat $c_{v}$ vs $T$ for several lattice sizes for the cases FCC and RCP.
The prominent peaks in $c_v$ for both cases hint at the existence of singularities, which is an expected feature in second order PM-FM transitions
in dipolar crystals. Both curves are compatible with a logarithmic divergence. Instead, the plot at $\Phi=0.426$ (see Fig.~\ref{heat}(c)) shows a
smooth curve with apparently no sign of singularity. This is the expected behavior in PM-SG transitions when there is strong frozen disorder.\cite{PADdilu, jpcm17}

\begin{figure}[!t]
\begin{center}
\includegraphics*[width=86mm]{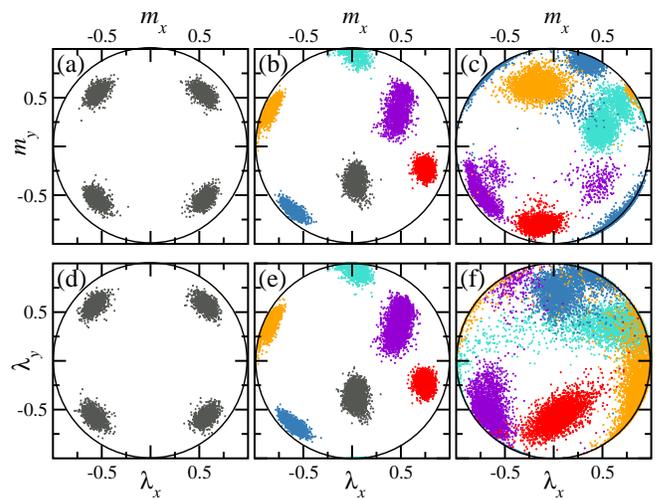}
\caption{(Color online) Upper row: thermal distribution of the $x$- and $y$-components of the normalized magnetization vector $\hat{m}$
  (provided that $m_z \ge 0 $). The samples are distinguished by the colors. (a) has been obtained in a FCC lattice; (b) and (c) at $\Phi=0.64$ and  
  $0.426$ respectively. Lower row: thermal distribution of the nematic director $\widehat{\lambda}$ (on the condition that
  $\lambda_z \ge 0 $.) (d) stands for the FCC lattice; (e) and (f) for $\Phi=0.64$, and $0.426$ respectively.
  All the distributions are for samples with $N=512$ particles at temperature $T=0.1$.}
  \label{supervector}
\end{center}
\end{figure}

Equilibrium distributions for the $x$- and $y$-components 
of the normalized magnetization vector $ \hat{m} \equiv \vec{m}/\Vert\vec{m}\Vert$
 and nematic director $\widehat{\lambda}$ at low temperature
offer a more precise picture of the type of order. They are shown in Fig.~\ref{supervector}. Panels (a) and (d) concern FCC systems and show that 
$\hat{m}$ and $\widehat{\lambda}$ are oriented along the four directions of the crystal, $(\pm 1,\pm 1, +1)$
in such a way that during the MC simulation, the entire configuration continuously flips between these directions.

On the contrary, all samples for the system at $\Phi=0.64$, each represented with a different color in the Figure, 
have only a single sample-dependent direction for both vectors, that fluctuate around them, see panels (b) and (e).
Only upon averaging over hundreds of samples, can we recover the expected
isotropy for those disordered systems. This behavior is reminiscent of the one encountered in the systems of Ising dipoles which have a fixed nematic director for each sample.

For small $\Phi$ we observe that the nematic director has no definite direction in many samples and that  the direction of $\hat{m}$ 
is not strongly coupled with that of  $\widehat{\lambda}$. Panels (c) and (f) of Fig.~\ref{supervector} show the distribution of the 
components of $\hat{m}$ and $\widehat{\lambda}$ for several samples at $\Phi=0.426$.

\begin{figure}[!b]
\begin{center}
\includegraphics*[width=84mm]{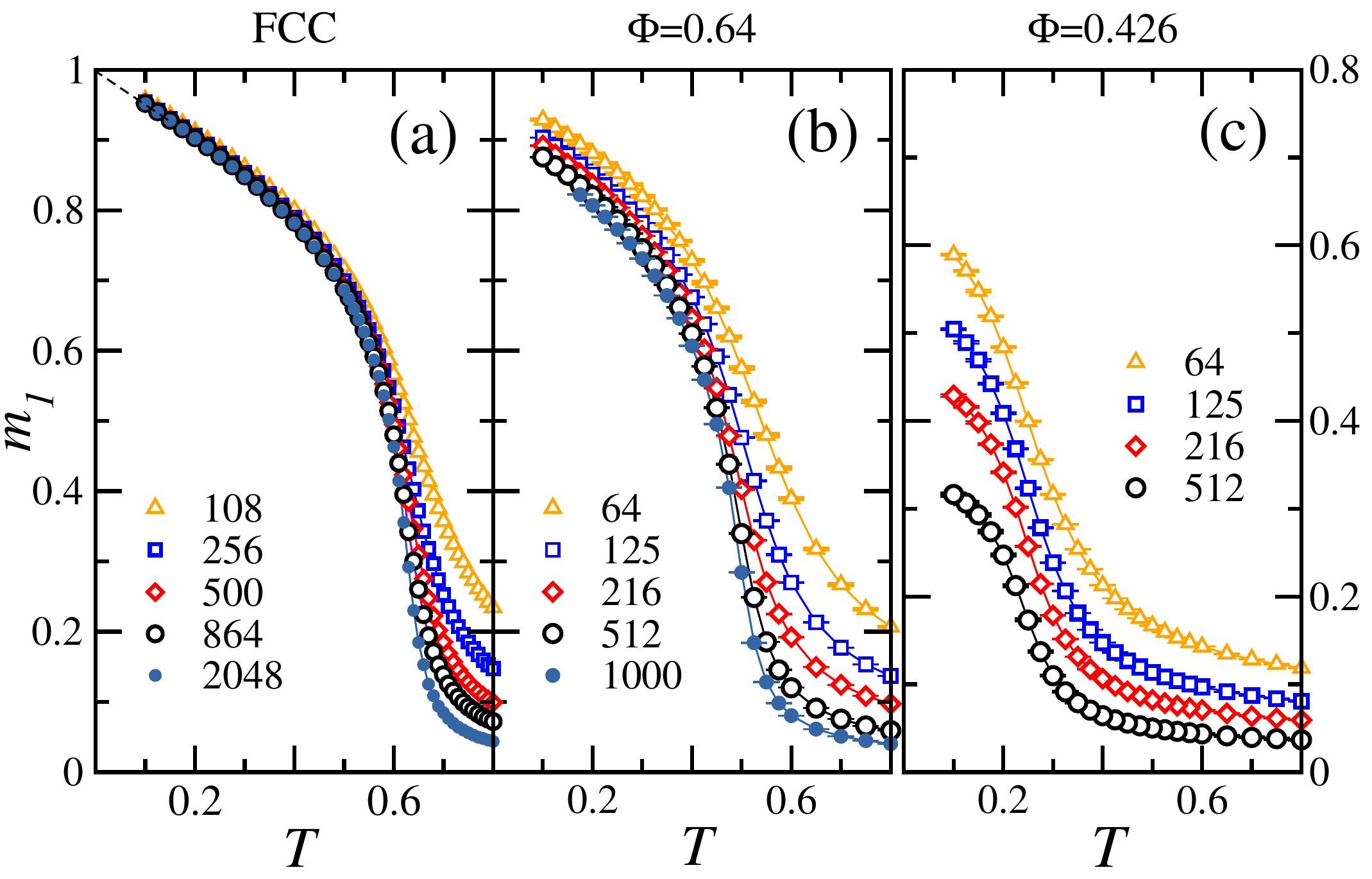}
\caption{ (Color online)  
(a) Plots of the magnetization $m_{1}$ vs $T$ for the FCC lattice and the number of dipoles $N$ indicated in the legend.
(b) Same as in (a) for RCP configurations ($\Phi=0.64$). (c) Same as in (b) but for $\Phi=0.426$. Lines in all panels are guides to the eye.}
\label{magne1}
\end{center}
\end{figure}

\subsection{FM order}
\label{FM}

The presence of strong long-range FM order is associated with a non--vanishing magnetization in the thermodynamic limit.
Fig.~\ref{magne1}(a) contains curves of $m_1$ vs temperature at various $N$ in a FCC crystal. They show that this is the case indeed:
$m_1$ is clearly independent of $N$ at low $T$ and tends to~1 for $T\to0$.

That conclusion differs for random packings with large $\Phi$, as shown by Fig.~\ref{magne1}(b) in the RCP limit. In this circumstance
$m_1$ does not saturate and clearly diminishes when $N$ grows for every $T$. Similar results are obtained for $\Phi \ge 0.5$.
The decay of $m_1$ is more obvious for less dense systems, and this makes evident the lack of any type of FM order, as shown in
Fig.~\ref{magne1}(c) for $\Phi=0.426$. This last finding agrees with the simulations of Refs\cite{ayton1,ayton2} for $\Phi=0.42$ in which
it was inferred that FM order is not present for all RDP.

\begin{figure}[!t]
\begin{center}
\includegraphics*[width=84mm]{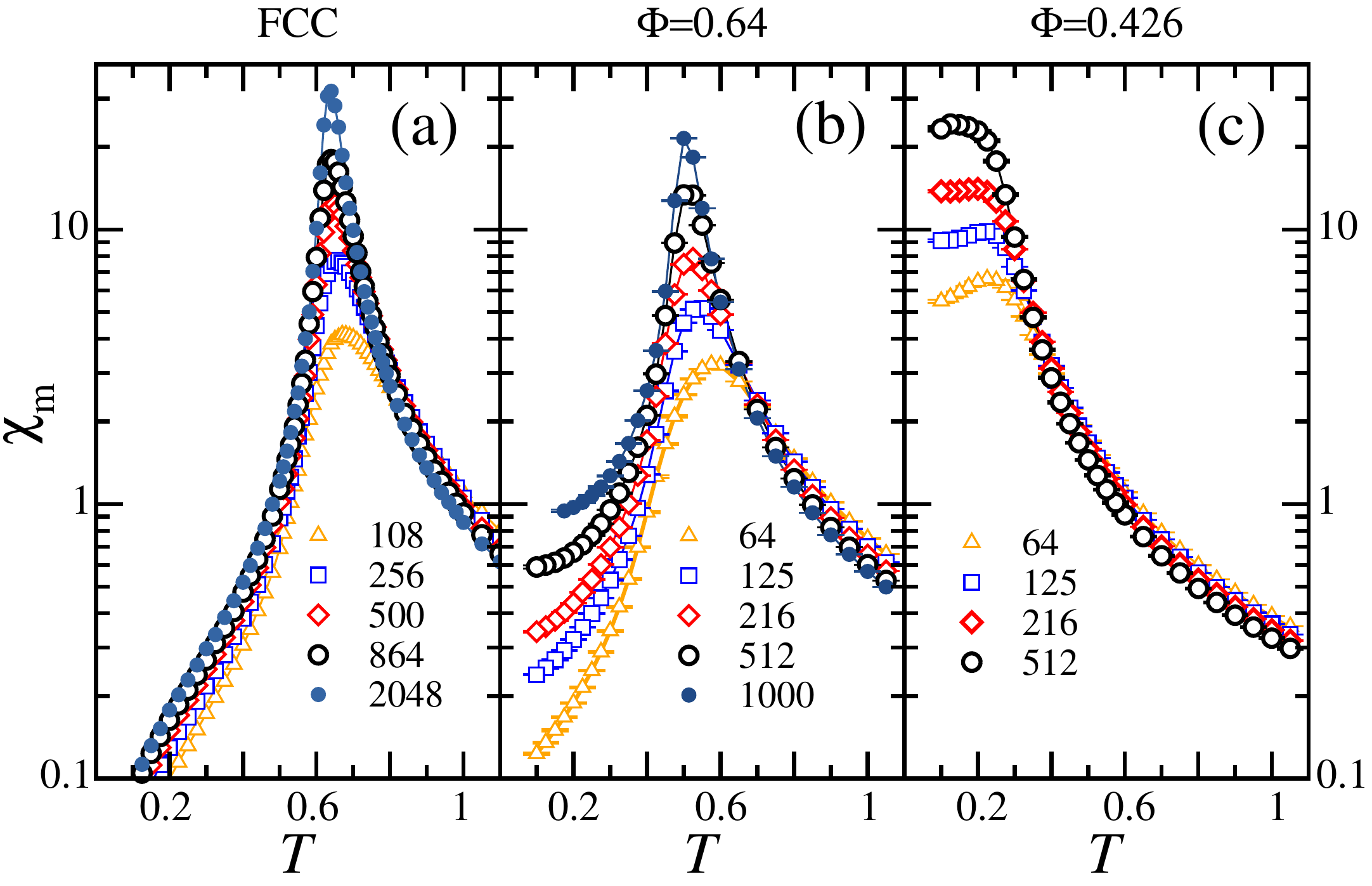}
\caption{
(Color online)
(a) Plots of the Log of the magnetic susceptibility $\chi_{m}$ vs $T$ on the FCC lattice and the number of dipoles $N$ indicated in the legend.
(b) Same as in (a) for RCP configurations ($\Phi=0.64$). (c) Same as in (b) but for $\Phi=0.426$. Solid lines in all panels are guides to the eye.}
\label{susc}
\end{center}
\end{figure}

\begin{figure}[!b]
\begin{center}
\includegraphics*[width=75mm]{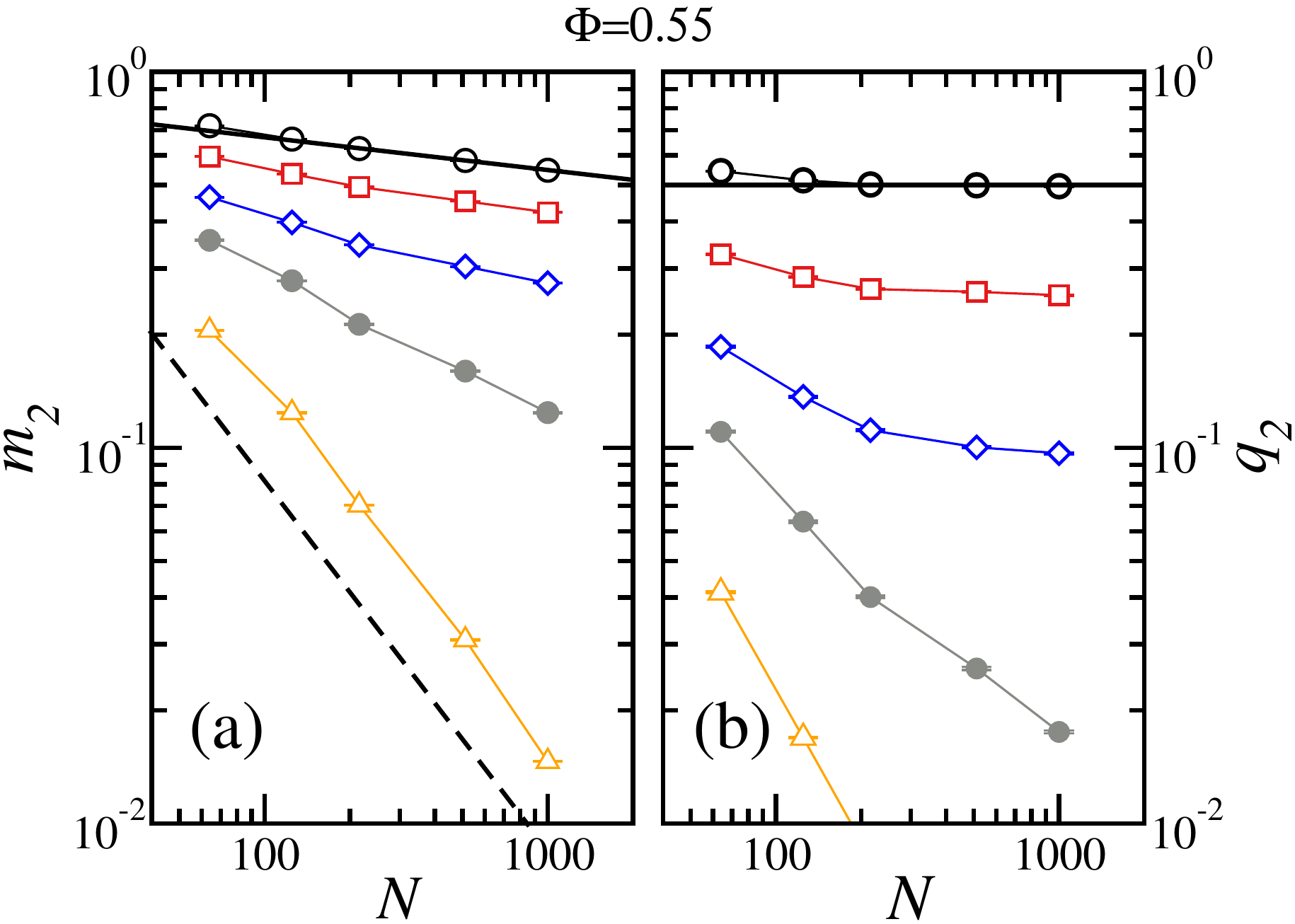}
\caption{ (Color online)
(a) Log-log plots of $m_{2}$ vs $N$ for $\Phi=0.55$. From top to bottom, 
$\smallcircle$,$\smallsquare$, $\smalldiamond$, $\smallblackcircle$m and $\smalltriangleup$ stand for $T=0.175, 0.275, 0.35, 0.4$, and $0.475$.
  (b) The same as in (a) for $q_{2}$. The thick solid lines in both panels show the power-law decay for the lowest temperature,  $T=0.175$. 
  The dashed line in panel (a) is the $N^{-1}$ decay expected for a PM phase. Lines connecting the data points are guides to the eye.}
\label{m2q2-55}
\end{center}
\end{figure}
 
Our results for $\Phi \ge 0.5$ point to a different interpretation. This is illustrated with the plots of the magnetic susceptibility $\chi_m$ vs $T$ of Fig.~\ref{susc}.
Panels (a) and (b) correspond to FCC and RCP respectively
and both exhibit a peak at a precise temperature $T_c$ that becomes sharper as $N$ grows, as it is expected
for PM-FM phase transitions of second order. In fact, our data is consistent in both cases with a power-law divergence of $\chi_{m}$ with  $N$.
Even more appealing is that $\chi_m$ diverge for all $T \le T_{c}$ in the RCP case, in contrast with the FCC case. This character of the plots for RCP is found throughout the region
$\Phi \ge 0.5$ and seems to indicate the existence of quasi-long-range (QLR)
order at $T \leq T_{c}$. The position of the peak of $\chi_m$ provides an
estimate of $T_c$ for different values of $\Phi$. We shall return to this when we will analyse the results for $B_m$.

The panel (c) in Fig.~\ref{susc} refers to data taken at $\Phi=0.462$. In this case we find no peak in spite of the fact that $\chi_m$ diverge at low temperatures. Both facts are typical signatures of SG phases.

\begin{figure}[!b]
\begin{center}
\includegraphics*[width=73mm]{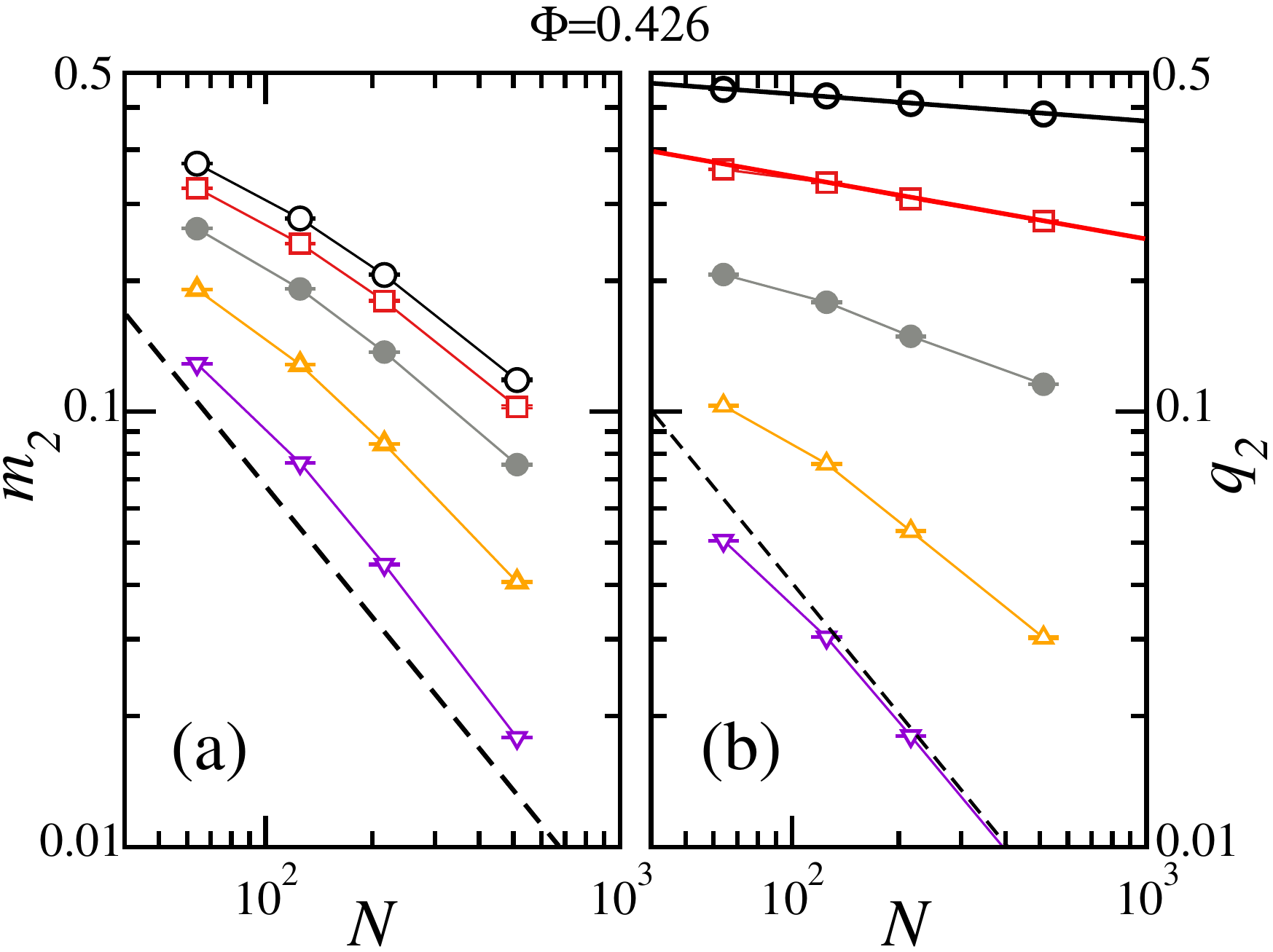}
\caption{ (Color online) 
(a) Log-log plots of $m_{2}$ vs $N$ for $\Phi=0.426$. From top to bottom, 
$\smallcircle$,$\smallsquare$, $\smallblackcircle$, $\smalltriangleup$, and $\smalltriangledown$ stand for $T=0.1, 0.15, 0.2, 0.25$, and $0.3$.
  (b) The same as in (a) for $q_{2}$. The thick solid lines shows the power-law decay for the lowest temperatures shown, $T=0.1$ and $0.15$. The dashed line in both 
   panels correspond to the $N^{-1}$ decay expected in a PM phase.}
\label{m2q2-426}
\end{center}
\end{figure}

To confirm the existence of QLR FM order for $\Phi \ge 0.5$ we studied the dependence of $m_2$ in the number $N$ of dipoles.
In Fig.~\ref{m2q2-55}(a) log-log plots of $m_2$ vs $N$ at various temperatures are shown for $\Phi =0.55$.
The transition temperature inferred from the position of the peak of $\chi_m$ is in this case $T_{c}=0.39(4)$.
Data from the Figure for $T \le T_{c}$ are consistent (at least for $N\ge216$) with a power-law decay of $m_2$ with $N^{-p}$ where $p$ is $T$-dependent.
The lattice sizes used in our work are not large enough to draw conclusions about the exponent $p$.
For temperatures slightly larger, the decay tends to be of the form $1/N$ which corresponds to a PM phase. 

Fig.~\ref{m2q2-426}(a) shows the analogous data
for $\Phi =0.426$. Now the curves of $m_{2}$ vs $N$ for all $T$ bend downwards with a slope that grows with $N$ and tends to the limit $1/N$. This is a clear signal of absence of FM order.

\begin{figure}[!t]
\begin{center}
\includegraphics*[width=88mm]{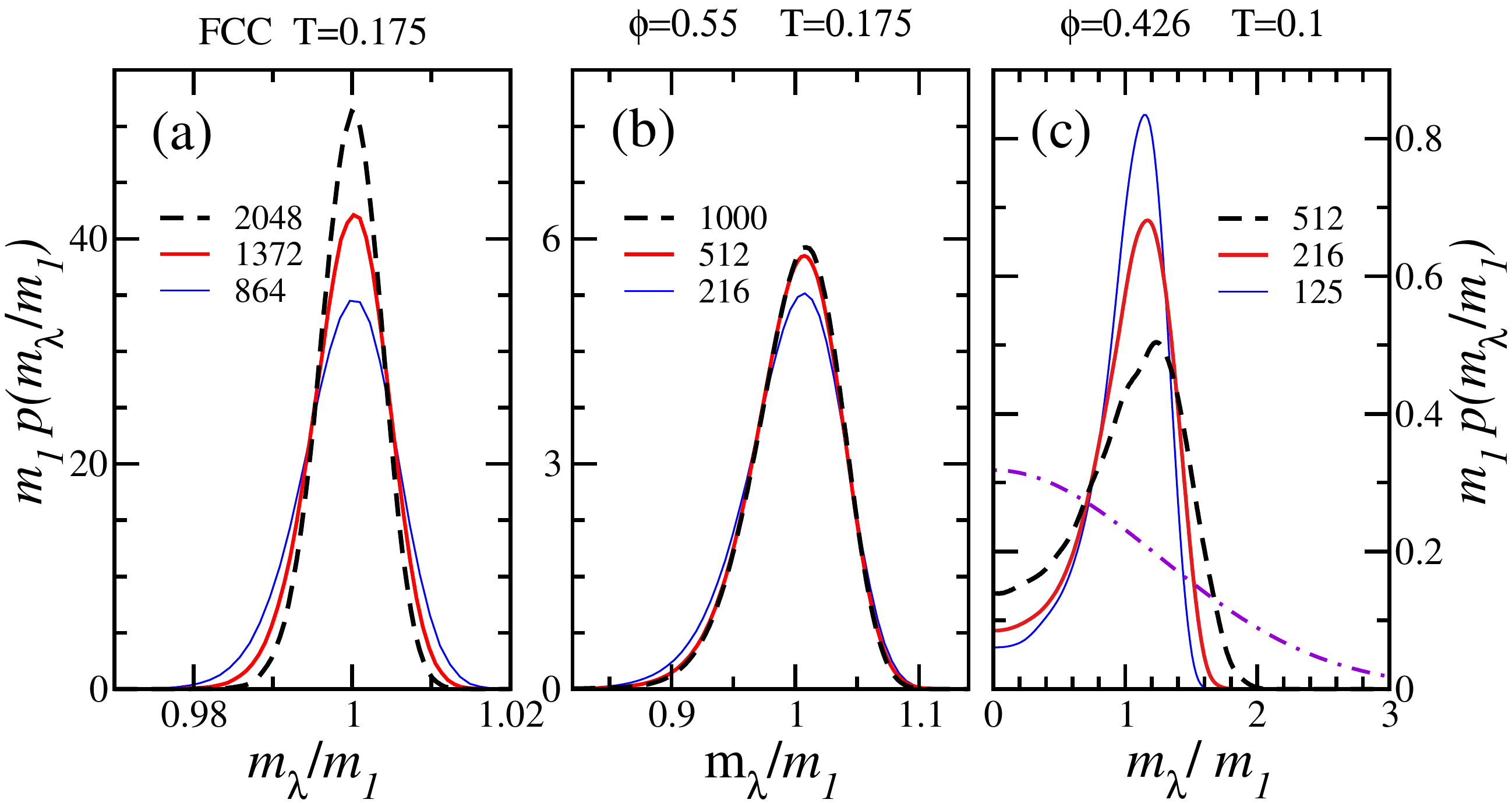}
\caption{ (Color online) 
 (Color online)
(a) Plots of the scaled probability distribution $m_1p(m_{\lambda}/m_{1})$ for the FCC lattice, 
temperature $T=0.175$, and  the number of dipoles $N$ indicated in the legend. 
(b) The same scaled distribution for systems with $\Phi=0.55$ and temperature $T=0.175$.
(c) The same distribution for systems with $\Phi=0.426$ and temperature $T=0.1$
 The dotted-dashed line stands for the Gaussian distribution of the PM phase in the $N \to \infty$ limit.}
\label{p-de-m}
\end{center}
\end{figure}

We can obtain additional information from the normalized distribution $p_{r} \equiv m_{1}~p(m_{\lambda}/m_{1})$ at low $T$.
If a marginal behavior exists for $\Phi \ge 0.5$ when $T \le T_{c}$, then $p_r$ is convenient because of its
independence of the size of the system, a typical trait near critical points.\cite{criti}
Figs.~\ref{p-de-m}(a-c) show $p_{r}$ for a handful of values of $N$ and for the FCC, $\Phi = 0.55$ and $0.426$ cases.
All distributions correspond to very low temperatures. For FCC the distribution becomes more peaked and narrower as $N$ grows, as it must be for a strong FM phase with
non-vanishing $m_1$. For $\Phi = 0.55$ the curves tend to coalesce as $N$ grows, another typical trait of criticality. All that indicates the presence of QLR FM order.

We end this description by interpreting the results for $\Phi = 0.426$. The related curves do not scale but broaden when $N$ grows. Only for sizes larger than those available in our simulations
(i.e. as long as the sizes of the magnetic domains shown in Figs.~\ref{portrait}(e,d) are less than the size of the system), and in the presence of FM order, these curves should tend to the Gaussian distribution
shown in the Figure. Thus, our results are consistent with the complete absence of FM order.
%
% (LO QUE HABIA): a fact that demonstrates the complete absence of FM order
%
%he quitado  lo de ''fact'', pues de hecho no vemos esa gaussiana en nuestras simulaciones ni de lejos:
%suponemos que deberíamos verla en sistemas de tamaño mucho mayor 
%
% (LO NUEVO)

%

\subsection{The PM-FM transition line}
\label{FMline}

The transition temperature $T_c$ can be extracted from the positions of the peaks in the plots for $c_v$ and $\chi_m$, and also
from analysing the Binder parameter $B_m$. The point is that, since the latter is scale invariant, the determination of $T_c$ from $B_m$ is
more precise. When there is long range strong order the value of $B_m$ tends to~1 when $T \le T_{c}$. However, the magnetic order in the
PM phase is short range and by the law of large numbers, we expect $B_{m}\rightarrow0$ when $N\to \infty$. Again, since $B_m$ is scale free,
it is independent of $N$ at the transition. Therefore, the plots of $B_m$ vs $T$ for various $N$ must cross at $T_c$ if the transition is of second order.
This is how the plots of the Binder cumulant allow to establish the value of $T_c$.

The results from the previous Section point out to the existence of a phase with QLR magnetic order for low $T$ when $\Phi \ge 0.5$.
That being so, $B_m$ should be independent of $N$ all over that phase and without reaching the value~1 when $N \to \infty$. Instead of crossing,
the plots of $B_m$ vs $T$
 should end up on top of each other forming one single curve for $T \le T_{c}$, at least for large enough $N$.\cite{balle}

\begin{figure}[!t]
\begin{center}
\includegraphics*[width=78mm]{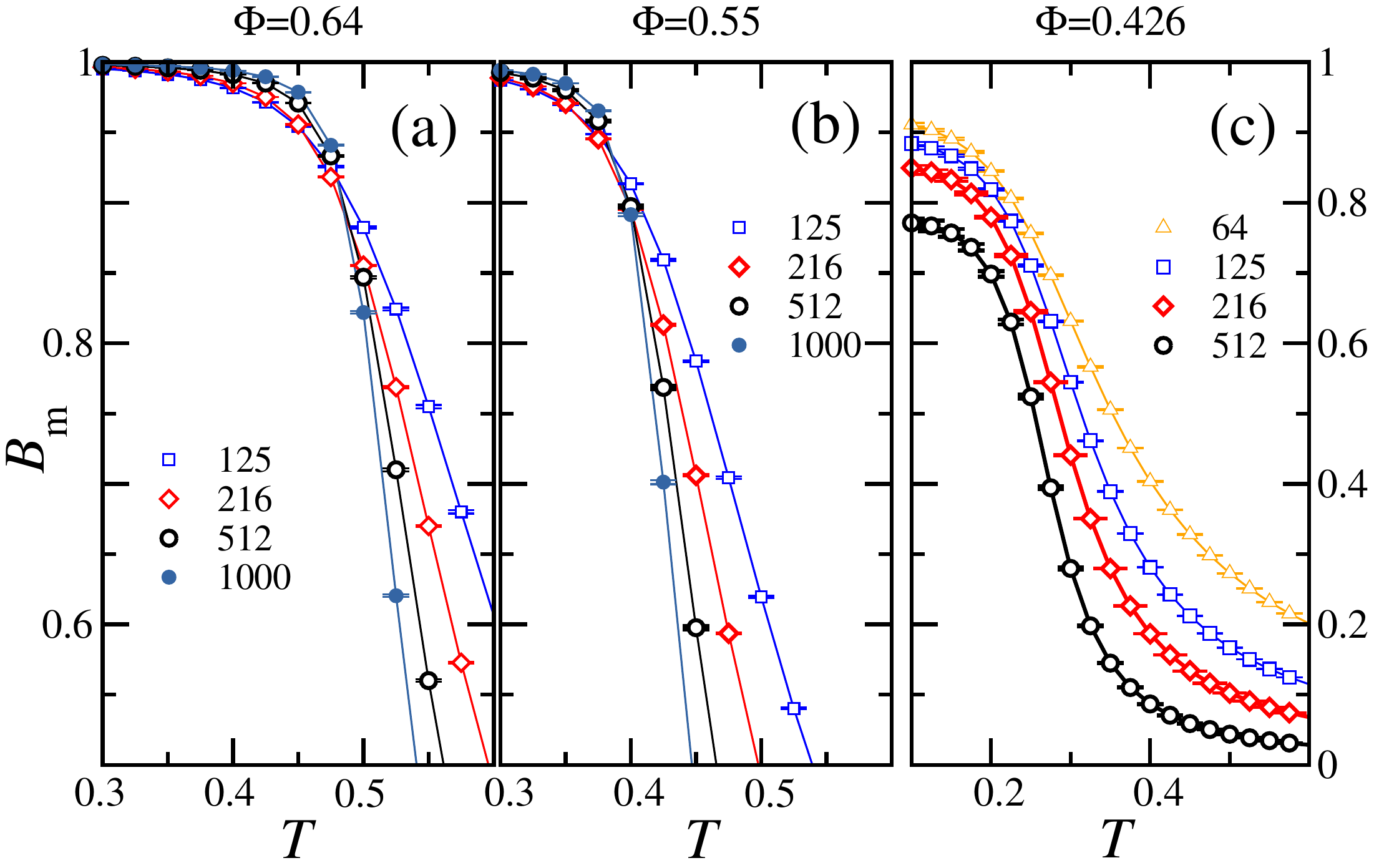} 
\caption{ 
(Color online) (a) Plots of the Binder cumulant $B_{m}$  vs  $T$ for the RCP case ($\Phi=0.64$) and the number of dipoles $N$ indicated in the legend.
(b) Same as in (a) for $\Phi=0.55$. (c) Same as in (b) for $\Phi=0.426$. Solid lines in all panels are guides to the eye.}
\label{binderM}
\end{center}
\end{figure}

\begin{figure}[!b]
\begin{center}
\includegraphics*[width=65mm]{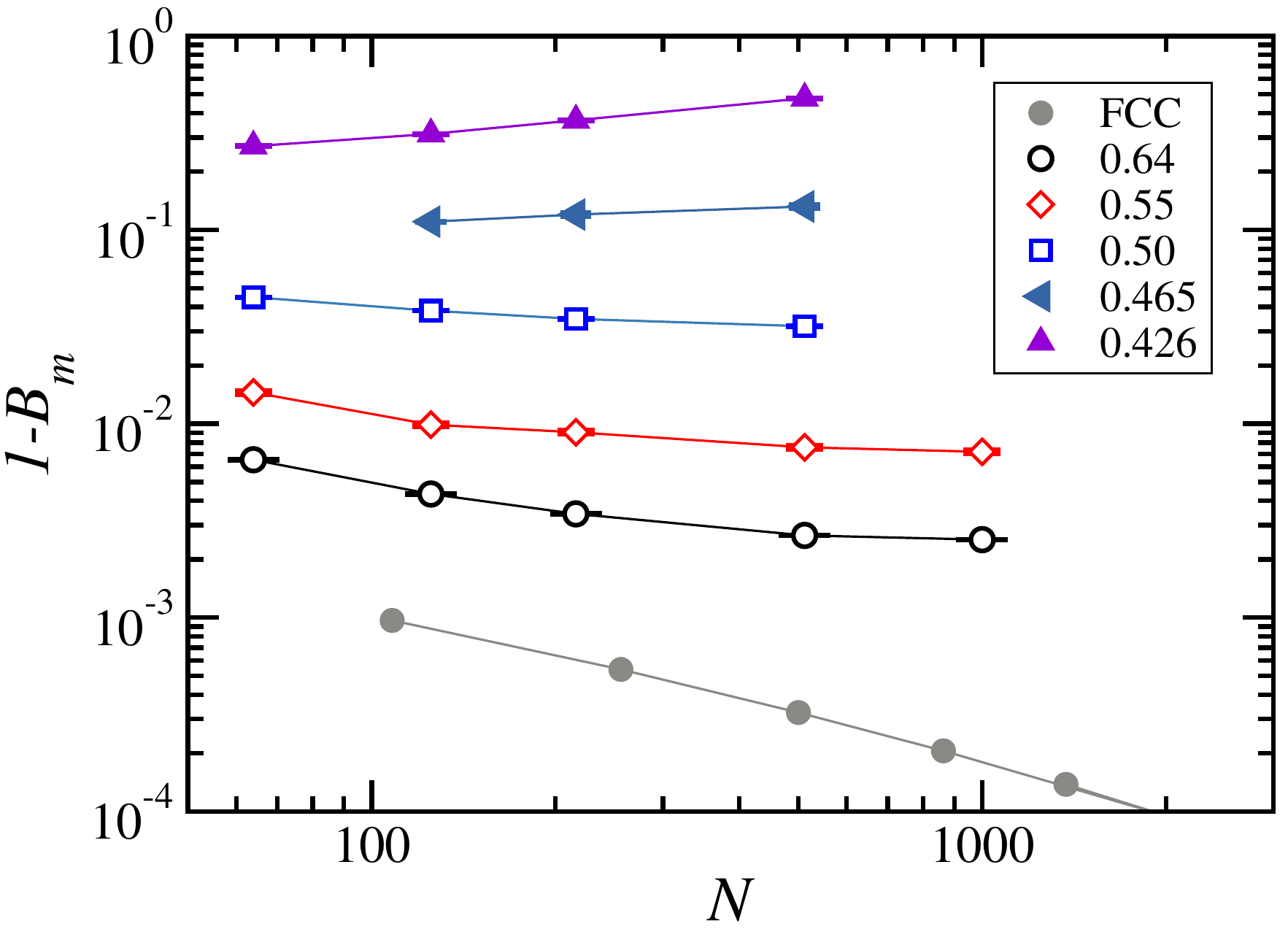}
\caption{ (Color online) 
  (a) Log-log plots of $1-B_{m}$ vs $N$ for temperature $T=0.175$  and the values of $\Phi$ indicated in the Figure.
  As stressed by the lines connecting the data points, $B_{m}$ does not saturate to $1$ in the thermodynamic limit
  for any random packing considered, in contrast with the FCC case. }
\label{binder-vs-N}
\end{center}
\end{figure}

In Fig.~\ref{binderM}(a,b) we show the plots of $B_{m}$ vs  $T$ for $\Phi=0.64$ and $0.55$. 
Although not shown, similar results  follow for $\Phi = 0.5$.
Then, we note that the curves cross when $N \ge 216$ in a rather precise point. This precision emphasizes the convenience of using the Binder cumulant for determining the
$\Phi$-dependence of $T_c$ and drawing the frontier between the PM and FM phases in the phase diagram, see Fig.~\ref{fases}.

The existence of such neat crossings may appear in contradiction with the possible existence of a marginal phase with QLR FM order.
To clear up all doubts, later we will verify that the Binder cumulant $B_m$ does not reach the value~1 at low temperatures in the thermodynamic limit.

The results for $\Phi < 0.5$ are qualitatively very different. We show in Fig.~\ref{binderM}(c) the plots $B_{m}$ vs  $T$ for $\Phi=0.426$. The value of $B_m$ diminishes as $N$ grows
for all $T$, revealing that there is no PM-FM transition. This suggests that no FM order exists for low values of $\Phi$. This hypothesis could be clinched if we were able to prove that
$B_{m} \to 0 $ as $N \to \infty$ even at low $T$. This test has been done in Fig.\ref{binder-vs-N} where the behavior of $1-B_m$ vs $N$ is studied at the lowest temperature we have
simulated, $T=0.175$, and for varying $\Phi$. All that is compared with data from FCC. In this latter case we observe a clear $B_m \to 1$ limit when $N\to\infty$, while for RDP with $\Phi \ge 0.5$
we see that $B_m$ tends to a value less than~1. This behavior is in accord with the existence of the above-mentioned marginal phase. On the contrary, for $\Phi < 0.5$
we observe that   $B_m$ tends to zero very slowly as $N$ grows. This indicates the absence of FM order.

\subsection{SG order}
\label{SG}

We have found no long-range strong FM order in RDP for any value of $\Phi$. In this Section we want to elucidate whether this lack of FM order may
give rise to SG order. An examination of the configurations shown in Fig.\ref{portrait} for $\Phi=0.5$ and $0.426$ reveals that the overlap between
different configurations of a single sample covers regions that are larger than the magnetic domains. This fact leads us to suspect that the SG order is
stronger than the FM order in both cases. To study the order of the SG phase, we analyse the overlap $q_{1}$ and the related quantities $B_{q}$ and $\xi_{L}/L$. 

\begin{figure}[!b]
\begin{center}
\includegraphics*[width=86mm]{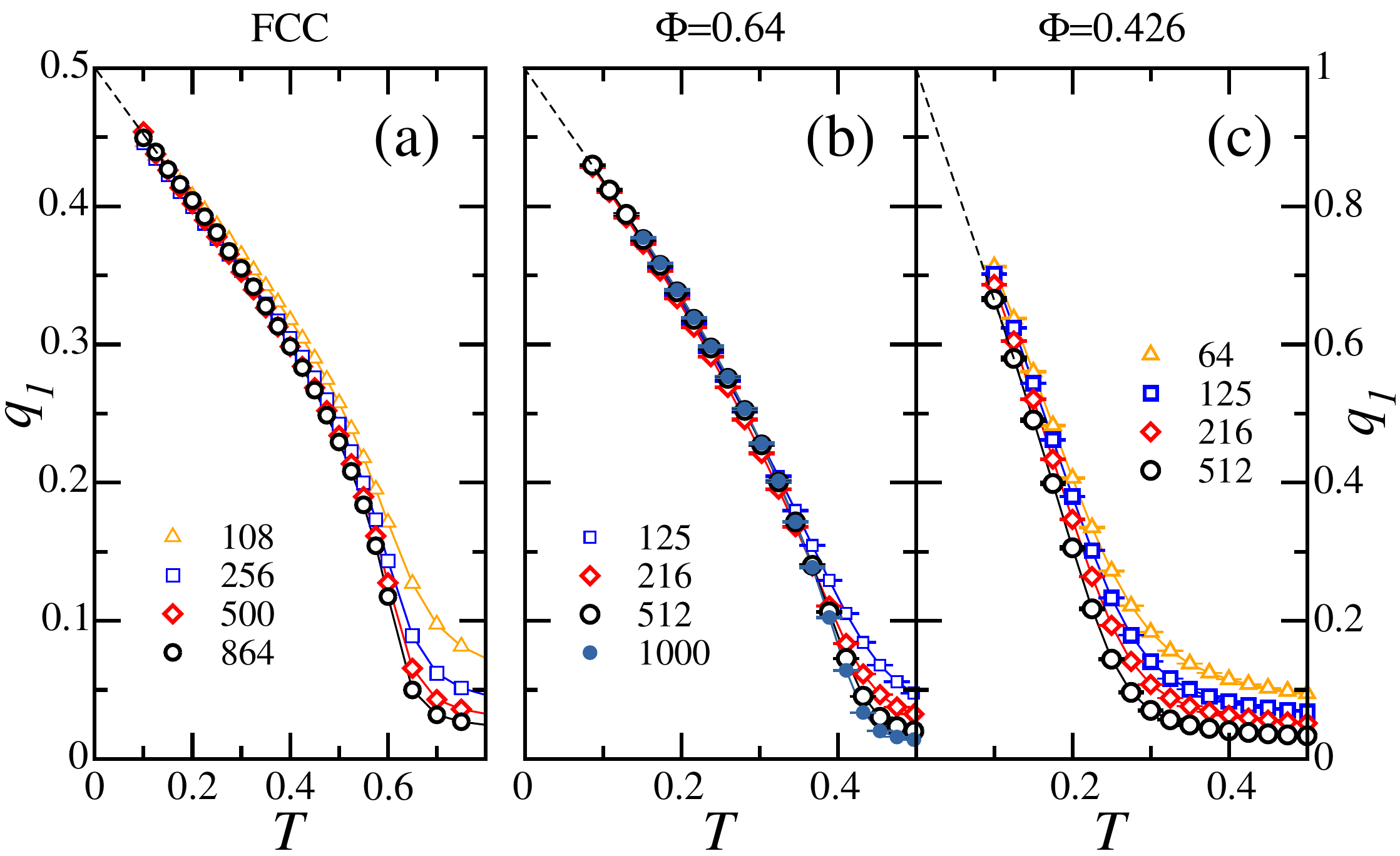}
\caption{(Color online)
(a) Plots of the scalar spin-glass overlap paramenter $q_{1}$  versus  $T$ for the RCP case ($\Phi=0.64$) and varying number of dipoles $N$, as indicated in the legend.
(b) Same as in (a) for $\Phi=0.55$. (c) Same as in (b) for $\Phi=0.426$. Solid lines in all panels are guides to the eye. Dashed lines are extrapolations for $T \to 0$.}
\label{q1}
\end{center}
\end{figure}

Fig.~\ref{q1} shows plots of $q_{1}$ vs $T$ for several $N$. The panels (a,b,c) correspond to the FCC, $\Phi=0.64$ and $0.426$ cases respectively. It is
illuminating to compare this Figure with its counterpart for $m_1$ in Fig.~\ref{magne1}. For FCC we find that $q_1$ does not go down as $N$ grows for
low temperatures. This is expected as neither does $m_1$ go down in this circumstance. We also note that $q_{1} \to {1}/{2}$ when $T\to 0$ in spite of
the fact that $m_{1}\to 1$. Recall that the vector $\vec{m}$ in FCC points equally in all crystalline directions $(\pm1,\pm1,+1)$, in such a way that the
TMC evolution jumps very easily from one to another.
The overlap $q$ is influenced by these global rotations and as a result its value is ${1}/{2}$.

For $\Phi =0.64$ we found that neither does $q_{1}$ become smaller as $N$ grows, like it occurs in the FCC case.
Actually this trend can be observed for all $\Phi\ge0.5$. What now happens is that $q_{1} \to 1$ for $T\to 0$, which
tells that the nematic order in each sample takes one single direction, in contrast to FCC. It is interesting to
compare the plots in Fig.~\ref{q1}(b) with those in Fig.~\ref{magne1}(b), in which $m_{1}$ goes down with $N$
for all temperatures. The different qualitative behaviors of the overlap and the magnetization are more clearly
seen in Fig.~\ref{m2q2-55} where the panels (a) and (b) show log-log plots of $m_2$ and $q_{2}$ vs $N$ for a
set of temperatures at $\Phi =0.55$. At low temperatures the plots of $q_{2}$ vs $N$ glaringly differ from a
power-law decay and bend upwards, hence $q_{2}$ does not vanish in the thermodynamic limit.
On the contrary, the plots for $m_2$ show the algebraic decay already noticed in the previous Section.
Finally, for $T\ge T_{c}$ we find that $q_{2}$ and $m_{2}$ go to zero if $N \to  \infty$, as it should occur in a FM phase.
Suming up, for $\Phi \ge 0.5$ we find a low temperature phase with QLR FM order and also strong SG order with $q\ne0$.

The behavior of the model is qualitatively different from what we have just explained if $\Phi < 0.5$. The plots in Fig.~\ref{q1}(c) for $\Phi=0.426$
show that for all temperatures $q_1$ decreases significantly as $N$ grows. To discover whether the overlap $q_1$ vanishes for $N\rightarrow \infty$
we constructed the log-log plots of $q_2$ vs $N$ in Fig.~\ref{m2q2-426}(b). The results for low $T$ are consistent with a $q_2 \sim 1/N^{p}$ functional
form with a $T$-dependent exponent $p$. Recall that the decay of $m_2$ was faster than a power-law and shows a tendency to a $1/N$ for large $N$ (see Fig.~\ref{m2q2-426}(a)),
as it would be expected for short-range FM order. All that is showing that there is a low temperature SG phase for $\Phi < 0.5$ with a marginal behavior and with short-range FM order.

\begin{figure}[!t]
\begin{center}
\includegraphics*[width=84mm]{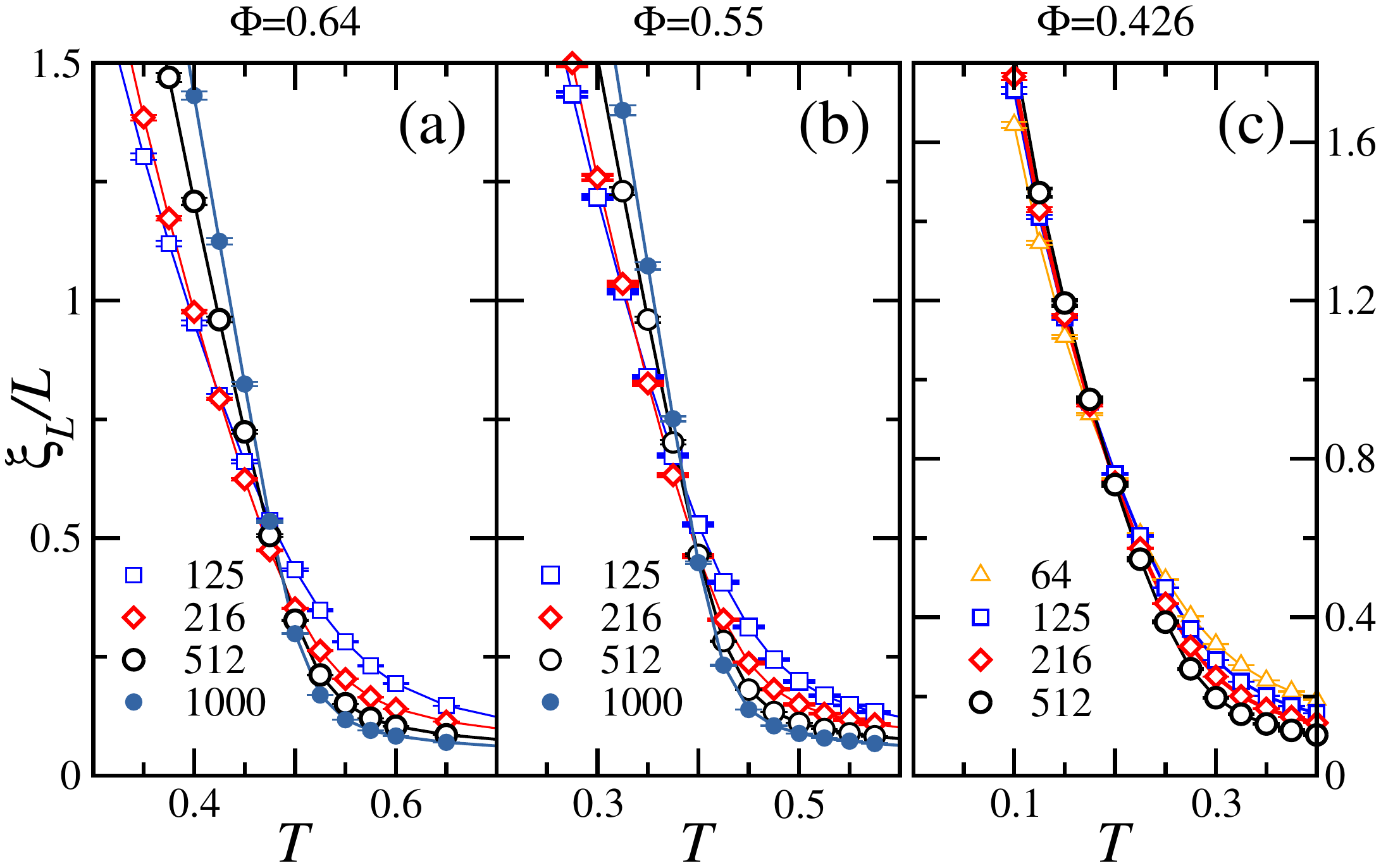}
\caption{(Color online) 
(a) Plots of the SG correlation length $\xi_{L}/L$  vs  $T$ for the RCP case ($\Phi=0.64$)  and varying number of dipoles $N$, as indicated in the legend.
(b) Same as in (a) for $\Phi=0.55$. (c) Same as in (b) for $\Phi=0.426$. The lines in all panels are guides to the eye.}
\label{longis}
\end{center}
\end{figure}

\subsection{The PM-SG transition line}
\label{SGline}

Next we wish to determine the temperature $T_{sg}$ at which the PM behavior yields a SG phase.
To this purpose we have measured the adimensional quantity $\xi_L/L$.\cite{longi,balle}. We stress that in a PM phase
(for which $\xi_L$ is a good approximation of the correlation length in SG), $\xi_L/L$ drops as $1/L$. Instead, when there
is strong long-range FM order, that is when $q \neq 0$, $\xi_L/L$ diverges as\cite{PADdilu} $L^{3/2}$.
Finally, for $T=T_{sg}$ the quantity $\xi_L/L$ becomes scale free and does not depend on $N$.
We expect that the plots of $\xi _L/L$ vs $T$ for various $N$ cross at $T_{sg}$ with a neat splay out of the curves above and below $T_{sg}$.
In the case of QLR SG order for $T < T_{sg}$ the several curves must coalesce for large enough $N$, since in this case $\xi_L/L$ does not diverge in the thermodynamic limit.

In Fig.~\ref{longis}(a,b) we present the above-described plots for $\Phi=0.64$ and $0.55$. We see that the plots for $N\ge216$ cross at a precise value $T_{sg}$ of the temperature.
This value defines the frontier between the regions where SG and the PM orders dominate. Within errors we find $T_{sg}$ equal to the Curie temperature $T_c$ obtained in the
previous Section from the plots for $B_m$. Equivalent results follow from the plots of $B_q$ vs $T$ apart from the fact that the crossing point for $\Phi=0.64$ occurs in a region
characterized by a dip that makes the determination of the transition temperature more difficult.\cite{korean, jpcm17}

\begin{figure}[!b]
\begin{center}
\includegraphics*[width=78mm]{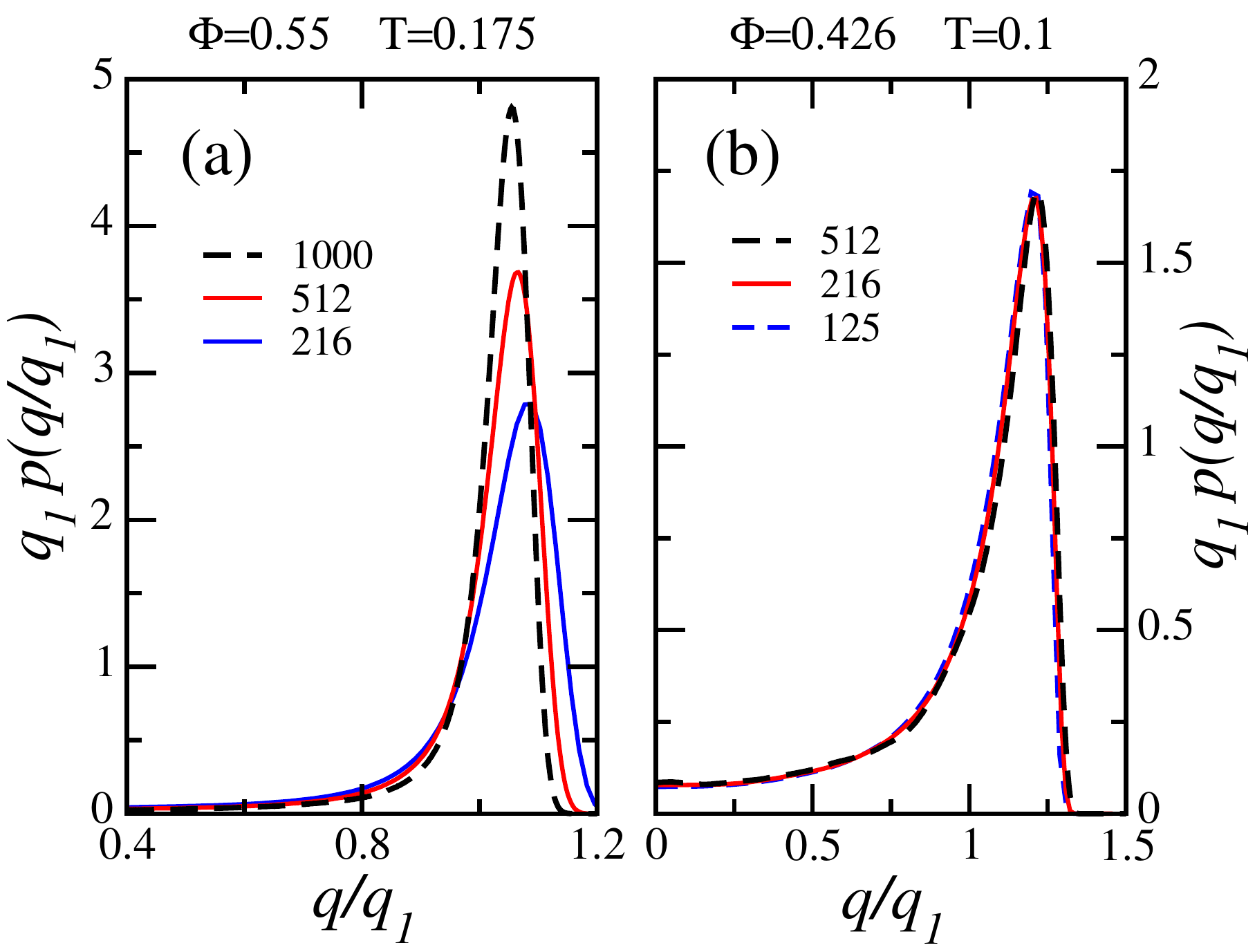}
\caption{ (Color online)
  (a) Plots of the scaled probability distribution $p(q/q_{1})$ for $\Phi=0.55$ and
  temperature $T=0.175$ and the number of dipoles $N$ indicated in the legend. 
  (b) The same distribution for $\Phi=0.426$ and temperature $T=0.1$.}
\label{p-de-q}
\end{center}
\end{figure}

\begin{figure}[!t]
\begin{center}
\includegraphics*[width=74mm]{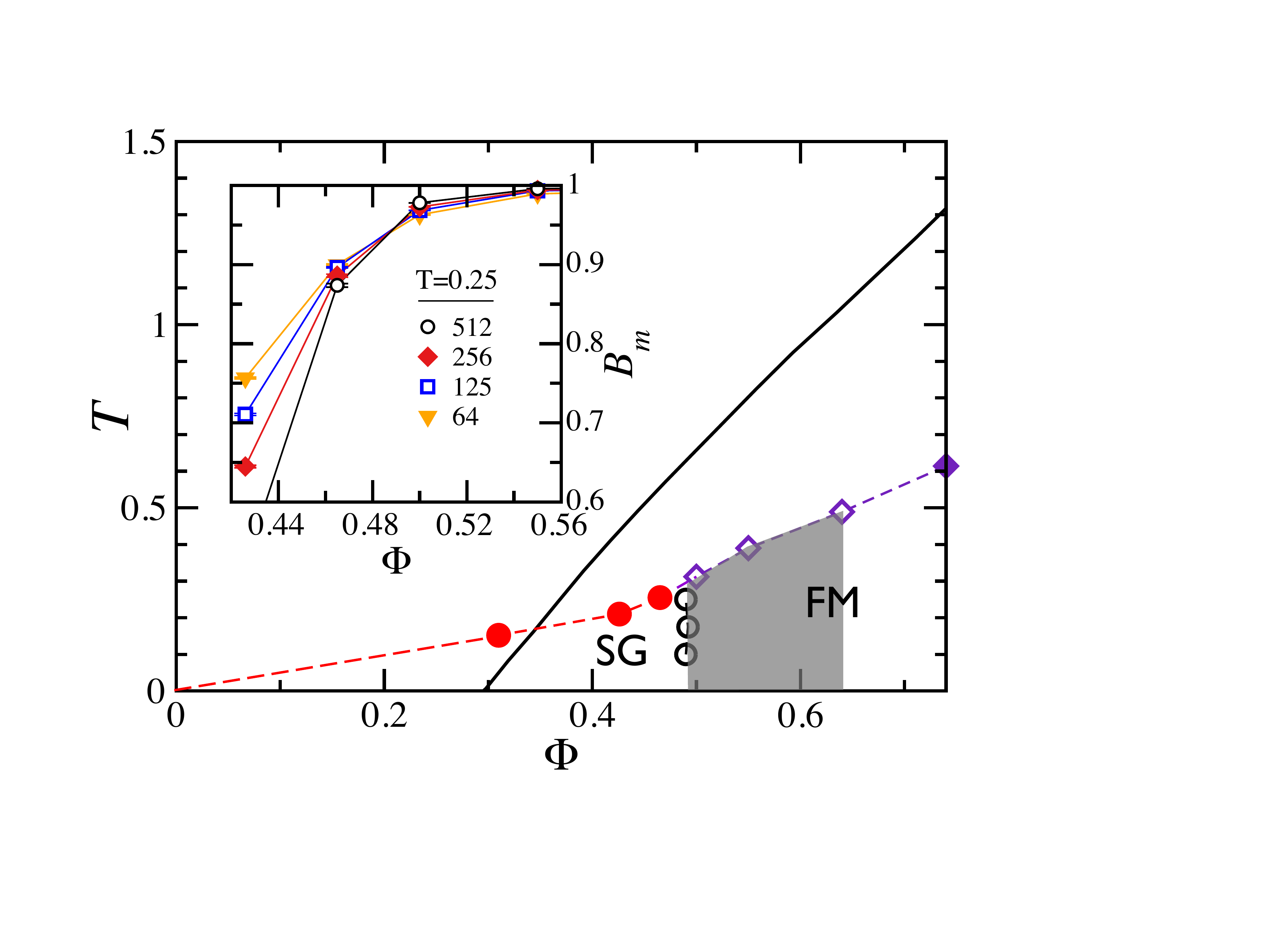}
\caption{
(Color online) 
 Phase diagram on the  $T-\Phi$ plane for the dipolar Heisenberg model. The symbols $\meddiamond$ delineate the  PM-FM transition for $\Phi \gtrsim 0.49$ (from $B_m$ vs $T$ plots.)
 $\blackdiamond$ stand for the FCC case.  Quasi long-range FM order with strong SG order has been found in the grey region.
 Symbols $\medblackcircle$ define the PM-SG transition (from $\xi_L$ vs $T$ plots.)
 $\medcircle$ are FM-SG transition points (from $B_m$ vs $\Phi$ plots.) The continuous line is the mean-field calculation of the FM-PM transition line. Dashed lines are guides to the eye.
 The inset contains plots of $B_{m}$  vs $\Phi$ for  $T=0.25$ and the number of dipoles $N$ indicated in the legend.}  
 \label{fases}
\end{center}
\end{figure}

Fig.~\ref{longis}(c) shows the curves of $\xi_L$  vs $T$ for $\Phi=0.426$. Actually we obtain qualitatively similar results for all $\Phi < 0.5$. Curves corresponding to different $N$ cross.
However, their splay out lessens as $N$ grows in the region $T<T_{sg}$. This makes the determination of $T_{sg}$ less accurate.
The plots for $B_q$ vs $T$ show a clear coincidence of all curves for $T<T_{sg}$ at all the sizes that we have analysed (not shown.)
This scenario is consistent with the presence of QLR SG order. The values of $T_{sg}$ are shown in the phase diagram of Fig.~\ref{fases}.
They define the region where the SG rules at $\Phi < 0.5$. Our results indicate that $T_{sg} \propto \Phi$ for dilute systems and that this phase extends until $\Phi\to0$.
This conclusions are in agreement with the results found for diluted systems of dipoles.\cite{stasiak, zhang-dilu}

To sustain  the evidence in favor of a strong SG order phase for $\Phi \ge 0.5$ and a marginal SG phase for $\Phi < 0.5 $ we have examined the thermal distribution $p(q/q_1)$
at low temperature in a similar fashion as it was done for $p(m_\lambda/m_1)$ in subsection~\ref{FM}. In Figs.\ref{p-de-q}(a-b) we present the normalized distribution 
$p_r \equiv q_1~p(q/q_1)$ at various values of $N$ for $\Phi = 0.55$ and $0.426$. This is a scaling function at criticality.
In the first case we observe that the distribution becomes sharp as $N$ grows, as it corresponds to a strong SG order with $q \neq 0$, see Fig.~\ref{p-de-q}(a). It is
worth noting that we find $p_r \to 0$ in the thermodynamic limit for small $q/q_1$, a fact that is in line with the droplet-model scenario for SG.\cite{RSB, bookstein}
On the other hand, for $\Phi < 0.5$ we find that the plots at different $N$ tend to coincide, in agreement with the above-mentioned marginal behavior.

In conclusion, the data for $\Phi \ge 0.5$ suggests the existence of strong SG order in the phase where we found QLR FM order. The results for $\Phi < 0.5$
indicate a SG phase for $T<T_{sg}(\Phi)$ where QLR SG order exists and FM order is absent. This SG phase is similar to the one found in systems of Ising
dipoles with strong structural disorder. In particular, this type of phases have been seen in textured systems with strong dilution,\cite{PADdilu, PADdilu2} as well as
in dense non-textured systems, that is with high disorder in the frozen directions of the Ising dipoles.\cite{jpcm17, alonso19, russier20}

\subsection{The FM-SG transition.}
\label{FMSG}

With the data gathered so far we can find the contours of the several FM and SG phases. To this end, we show plots of $B_{m}$ vs $\Phi$ for various $N$ along
the isothermals with $T$ below the PM boundary. We must not forget that $B_{m}$ goes down when $N$ increases in the SG phase, while for the marginal FM phase $B_m$ increases with $N$
with a limiting value less than~1. For that reason we suppose that the plots of $B_{m}$ vs $\Phi$ will cross at a transition point $\Phi_{tr}(T)$. This is indeed the case, as shown in the inset of
Fig.~\ref{fases} for $T=0.25$. The transition points obtained in this way are shown in the main picture of Fig.~\ref{fases}. The accuracy is poor since we have few available values of $\Phi$ and the
lattice sizes are not very large. Within these limitations, we find that this boundary line is vertical and placed   at $\Phi = 0.49(1)$. We find no signs of reentrances.

\section{CONCLUSIONS }
\label{conclusion}

We have studied by Monte Carlo simulations the role played by positional frozen disorder in the collective behavior 
of disordered dense packings  of identical  magnetic nanoparticles (NP) 
that behave as Heisenberg dipoles. These dipoles are free to rotate and deprived of local anisotropies.
The amount of structural disorder has been assessed by the volume occupancy fraction $\Phi$.

Although actual single domain NP cannot be free of local
anisotropies, the present study is relevant
for the field of NP because it shows the effect of the structural disorder 
on the phase diagram of systems of NP  in the small anisotropy limit.
It must be interpreted in the same way as dipolar
Ising models can be used to model the several facets of the strong anisotropy limit.

The results allow to obtain the phase diagram on the temperature-$\Phi $ plane (see Fig.~\ref{fases}.)
Concretely we have studied the magnetization $m_{\lambda}$, the scalar spin-glass overlap parameter $q$,
and related fluctuations. The Binder parameters for $m_{\lambda}$ and $q$ and the SG correlation length
offer the opportunity of determining the extent of the regions with ordered low-temperature phases.  

For random dense packings with $\Phi \gtrsim 0.49$ (including the limiting random-close-packed case) we find a well defined second 
order transition line that separates a ferromagnetic (FM) phase from a high-temperature paramagnetic (PM) phase.
In contrast with the strong  FM order found for face-centered cubic (FCC) lattices, the FM 
phase for random dense packings exhibit signatures of quasi-long-range FM order and, 
at the same time, signatures of strong long-range spin-glass (SG) order with a non-vanishing overlap
parameter $q$ in the thermodynamic limit. A similar phase has been found for 
the random anisotropy Heisenberg magnet with short-ranged interactions,\cite{itakura} 
and for non-textured FCC systems of dipoles with low but not negligible anisotropy.\cite{enviado}
% falta completar la cita ''enviado''
%

For $\Phi \lesssim 0.49$,  the marginal FM order
disappears giving rise to a dipolar SG phase with quasi-long-range order. This marginal SG phase
is qualitatively similar to the one found in several systems of Ising dipoles with strong structural disorder.
Our results for relatively small $\Phi$ suggest that the SG phase extends to $\Phi \to 0$ with a
transition temperature $T_{sg} \propto \Phi $.

\section*{Acknowledgements}
We thank the Centro de Supercomputaci\'on y Bioinform\'atica  at University of M\'alaga, 
Institute Carlos I at University of Granada and Cineca (Italy) for their generous allocations
of computer time.  
This work was granted an access to the HPC resources of
CINES under the allocations 2019-A0060906180 and 2020-A0080906180
made by GENCI, CINES, France.
Finantial support from grants FIS2017-84256-P (FEDER funds) from
the Spanish Ministry and the Agencia Espa{\~n}ola de
Investigaci{\'o}n (AEI), SOMM17/6105/UGR from Consejer\'{\i}a de Conocimiento,
Investigaci\'on y Universidad, Junta de Andaluc\'{\i}a and European Regional Development Fund (ERDF),  
and Iniziativa Specifica NPQCD from INFN (Italy) are grafefully aknowledged.

\end{document}